\documentclass[preprint2]{emulateapj}
 \usepackage{graphicx,makecell,multirow,color,float}
 \usepackage{url}
 \usepackage{enumerate}
 \usepackage[flushleft]{threeparttable}

\def\Lya{Ly$\alpha$}
\def\EWLya{$\mathrm{EW_{Ly\alpha}}$}
\def\EWOIII{$\mathrm{EW_{[O\,III]}}$}
\def\dvLya{$\Delta v_{\mathrm{Ly\alpha}}$}

\def\Hb{H$\beta$}
\def\OII{[O\,{\sc ii}]}
\def\OIII{[O\,{\sc iii}]}
\def\Othreetwo{$\rm O_{32}$}

\def\HI{H\,{\sc i}}
\def\HII{H\,{\sc ii}}

\def\fesc{$f_{\rm esc}$}
\def\fescLya{$f_{\rm esc}^{\rm Ly\alpha}$}
\def\fescLyc{$f_{\rm esc}^{\rm LyC}$}
\def\fescrel{${f}_{\mathrm{esc,rel}}$}
\def\xiion{$\xi_{\rm ion}$}
\def\MUV{$\rm M_{UV}$}

\def\fescLya{$f_{\rm esc}^{{\rm Ly}\alpha}$}

\def\HII{H\,{\sc ii}}

\usepackage{color}

\shorttitle{LACES: Ionizing Radiation at $z=3.1$}
\shortauthors{Fletcher et al.}

\begin{document}

\title{The Lyman Continuum Escape Survey: Ionizing Radiation from [O III]-Strong Sources at a Redshift of 3.1}

\author{Thomas J. Fletcher\altaffilmark{1},
Mengtao Tang\altaffilmark{2},
Brant E. Robertson\altaffilmark{3,4},
Kimihiko Nakajima\altaffilmark{1,5,6},
Richard S. Ellis\altaffilmark{1},
Daniel P. Stark\altaffilmark{2},
Akio Inoue\altaffilmark{7}
}

\altaffiltext{1}{Department of Physics and Astronomy, University College London, Gower Street, London WC1E 6BT, UK}
\altaffiltext{2}{Steward Observatory, University of Arizona, 933 N Cherry Ave, Tucson, AZ 85721, USA}
\altaffiltext{3}{Institute for Advanced Study, 1 Einstein Drive, Princeton, NJ 08540, USA}
\altaffiltext{4}{Department of Astronomy \& Astrophysics, University of California, Santa Cruz, 1156 High St, Santa Cruz CA 95064, USA}
\altaffiltext{5}{European Southern Observatory (ESO), Karl-Schwarzschild-Strasse 2, 85748 Garching, Germany}
\altaffiltext{6}{National Astronomical Observatory of Japan, 2-21-1 Osawa, Mitaka, Tokyo 181-8588, Japan}
\altaffiltext{7}{Osaka Sangyo University, 3-1-1 Nakagaito, Daito, Osaka 574-8530 Japan}


\begin{abstract}
We present results from the LymAn Continuum Escape Survey (LACES), a Hubble Space Telescope (HST) program designed to characterize the ionizing radiation emerging from a sample of Lyman alpha emitting galaxies at redshift $z\simeq 3.1$. As many show intense \OIII{} emission characteristic of $z>6.5$ star-forming galaxies, they may represent valuable low redshift analogs of galaxies in the reionization era. Using HST Wide Field Camera 3 / UVIS $F336W$ to image Lyman continuum emission, we investigate the escape fraction of ionizing photons in this sample. For 61 sources, of which 77\% are spectroscopically confirmed and 53 have measures of \OIII{} emission, we detect Lyman continuum leakage in 20\%, a rate significantly higher than is seen in individual continuum-selected Lyman break galaxies. We estimate there is a 98\% probability that $\leq 2$ of our detections could be affected by foreground contamination. Fitting multi-band spectral energy distributions (SEDs) to take account of the varying stellar populations, dust extinctions and metallicities, we derive individual Lyman continuum escape fractions corrected for foreground intergalactic absorption. We find escape fractions of 15 to 60\% for individual objects, and infer an average 20\% escape fraction by fitting composite SEDs for our detected samples. Surprisingly however, even a deep stack of those sources with no individual $F336W$ detections provides a stringent upper limit on the average escape fraction of less than 0.5\%. We examine various correlations with source properties and discuss the implications in the context of the popular picture that cosmic reionization is driven by such compact, low metallicity star-forming galaxies.

\end{abstract}

\keywords{galaxies: distances and redshifts, evolution, formation, star formation -- cosmology
: early universe --  infrared: galaxies}


\section{Introduction} \label{sec:intro}

Deep imaging with the WFC3/IR camera on-board HST has dramatically expanded our redshift horizon, making it practical to address two long-standing cosmological questions: (i) when did the Universe transition from a neutral to an ionized state, and (ii) were early star-forming galaxies responsible for this cosmic reionization? Multi-color HST/Spitzer imaging in the Ultra Deep Field \citep[UDF;][]{beckwith2006a,ellis2013udf,koekemoer2013a,illingworth2013a} and the CANDELS fields \citep{grogin2011a,koekemoer2011a,oesch2013candels,bouwens2015uv}, together with complementary studies undertaken through the CLASH \citep{bradley2014} and Frontier Field \citep{mcleod2015,lotz2017a} lensing clusters, have delivered several hundred $z>7$ Lyman break galaxy (LBG) candidates providing the first convincing description of the abundance and luminosity distribution \citep{mclure2013,atek2015,finkelstein2015a,bouwens2015uv,livermore2017,ishigaki2018} of early star-forming galaxies to $z\simeq 10$ (see \citealt{stark2016} for a review).

The optical depth, $\tau$, of electron scattering to the cosmic microwave background measured by the Planck consortium (Planck 2015) constrains the redshift window over which reionization occurred. \cite{brant2015constraints} demonstrated how the demographics of star-forming galaxies determined by HST can be reconciled with this value in terms of a reionization history over the redshift range $6 \lesssim z \lesssim 12$ given some significant assumptions about the ionizing capability of the typical, most abundant, low luminosity sources. The key assumptions relate to (i) the UV radiation emerging from their stellar populations, defined by \cite{robertson2013} in terms of \xiion{}, the number of Lyman continuum (LyC) photons produced per UV (1500 $\mbox{\normalfont{\AA}}$) luminosity, and (ii) the fraction \fesc{} of such LyC photons that can escape absorption within the galaxy and its immediate vicinity. The quantity \xiion{} cannot be determined from broad-band photometry alone \citep{robertson2013} and is best constrained from Balmer line emission using recombination physics with a weak dependence on \fesc{} \citep{bouwens2016}. Until JWST is launched, the relevant lines are beyond reach of ground-based spectrographs at high redshift. Likewise the opacity of the intergalactic medium at UV wavelengths becomes too great beyond $z\simeq 4$ to determine \fesc{} from deep HST imaging below the Lyman limit (e.g., \citealt{shapley2006}). Therefore, neither of these quantities can be constrained for galaxies in the reionization era with current facilities, yet they collectively comprise the primary uncertainty in claims that reionization is driven by star-forming galaxies. The situation is particularly critical for \fesc{} since \cite{robertson2013, brant2015constraints} argued a mean value of $10-20\%$ is required for galaxies to reionize the Universe, whereas studies at redshifts where LyC photons can be directly detected frequently yield upper limits of $f_{esc}\lesssim5\%$ (e.g., \citealt{siana2015,mostardi2015}).

In promoting the view that early star-forming galaxies reionized the Universe, many workers have speculated that the both the intensity of the ionizing radiation (effectively \xiion{}) and the porosity of neutral gas in the circumgalactic medium (i.e. \fesc{}) increase with redshift, particularly for compact, intensely star-forming systems \citep{inoue2006,kuhlen2012,robertson2013,finkelstein2015a}.
Indeed, the strength of nebular emission (e.g., [O III] 5007 $\mbox{\normalfont{\AA}}$), whether measured directly from near-infrared spectroscopy \citep{schenker2013}, or inferred indirectly from the excess flux in Spitzer photometry \citep{labbe2013, smit2014, smit2015}, does apparently increase with redshift. Surprisingly, some of the most luminous LBGs with large \EWOIII{} at $z>7$ \citep{guido2016} also reveal \Lya{} in emission \citep{oesch2015,zitrin2015,laporte2017,stark2017}, `bucking the trend' established for less luminous systems. This correlation may imply that sources with large \EWOIII{} also have a high value of \fesc{}, thereby creating early ionized bubbles that permit \Lya{} photons to emerge \citep{stark2017}.

The inter-dependence of large \EWOIII{}, a higher than average value of \xiion{}, and the leakage of LyC photons was first evaluated in the context of photoionization models by \cite{nakajima2014}. Compiling literature data, they found an interesting correlation between the emission line ratio \OIII{}/\OII{} (hereafter \Othreetwo) and \fesc{}, which they claimed arises when \HII{} regions are ``density-bound" and some LyC leakage occurs. This picture contrasts with typical ``ionization-bound" \HII{} regions where LyC photons are fully absorbed within the radius of the associated Stromgren sphere.
The conjecture has received further support by the recent detections of significant LyC radiation from nearby intense \OIII{} emitters \citep{izotov2016nature,izotov2016, izotov2018}. The most extreme \Othreetwo{} sources in the \cite{nakajima2014} study were narrow-band selected Lyman alpha emitters (LAEs) whose observed properties are in many respects very similar to the dominant population of star-forming galaxies during reionization. Subsequently, through near-infrared and optical spectroscopy, \cite{nakajima2016,nakajima2018} provided further evidence that such LAEs have higher values of \xiion{} than continuum-selected LBGs \citep{shivaei2018},
and \citet{tang2018a} showed that \xiion scales with \OIII{} EW, reaching very large values in the most intense line emitters..

The most practical route to determine whether early galaxies reionized the IGM is to undertake a detailed study of analogs of this population at the highest redshift where direct measures of \xiion{} and \fesc{} are possible. With a representative sample of such analogs it may be possible to verify the inferred correlation between the \OIII{} emission and \fesc{}, as well as to determine the fraction of sources whose ionizing output (as defined by \xiion{} and \fesc{}) would be sufficient if projected, into the $z>7$ population, to sustain reionization. Intermediate redshift LAEs possibly represent the most valuable low redshift analogs of the population of compact, low mass, intensely-star forming galaxies that dominate the reionization era. Taking advantage of a large area narrow-band selected sample of spectroscopically-confirmed LAEs in the SSA22 field \citep{hayashino2004, matsuda2005, yamada2012, nakajima2016, nakajima2018}, the LACES project (the LymAn Continuum Escape Survey) aims to study these sources in detail and in particular to examine their LyC leakage via Hubble Space Telescope (HST) broad-band imaging below the Lyman limit. A key question our survey can address is whether intense \OIII{} emission seen in many LAEs is associated with an increased \fesc{} as conjectured originally by \cite{nakajima2014}. Deep UV imaging (and hence measures of \fesc{}) is presented for an unique and representative sample of $z\simeq 3.1$ LAE analogs for which the associated measures of Ly$\alpha$ and \OIII{} are already available from Keck and VLT spectroscopy.

A plan of the paper follows. In Section \ref{sec:data} we introduce our sample which is based on HST imaging in 3 WFC3 fields spanning the area for which we have extensive ground-based optical and near-infrared spectroscopy. In this section we discuss the relevant imaging and spectroscopic data and their processing. In Section \ref{sec:candidates} we examine the new deep $F336W$ images and devise a procedure for determining the presence of LyC leakage on a case by case basis in our sample, as well as the combined flux from those sources without individual detections. In Section \ref{sec:analysis} we define a path for deriving the measured \fesc{} or limits on its value from the individual $F336W$ fluxes, noting the dependences on the assumed form of the UV continuum as probed by independent spectroscopic measures of \xiion{}. In Section \ref{sec:results} we then correlate these measures with various source properties measured either observationally or derived from our model-dependent analyses. In Section \ref{sec:discussion}, we discuss these correlations in the context of whether such \OIII{}-intense sources are likely prominent agents of cosmic reionization.

Throughout this paper, we adopt a concordance cosmology with $\Omega_{\Lambda}=0.7$, $\Omega_{m}=0.3$ and $H_{0}=70\ \mathrm{km}\ \mathrm{s}^{-1}\ \mathrm{Mpc}^{-1}$. All magnitudes are given in the AB system \citep{okegunn1983}. When we refer to \fesc{} we mean the absolute escape fraction of LyC photons unless specified otherwise.


\section{Data}\label{sec:data}

As discussed in \cite{nakajima2016}, our target sample is drawn from a Subaru imaging survey that identified $z \simeq 3.1$ Lyman alpha emitters (LAEs) in the SSA22 field \citep{hayashino2004,matsuda2005,yamada2012} via their photometric excess in a narrow band filter at 497 nm. In addition to initial spectroscopy to confirm their identity, we have undertaken a systematic campaign using optical spectrographs on Keck and the VLT to study their rest-frame UV emission lines \citep{nakajima2018} and near-infrared spectroscopy with Keck's MOSFIRE to examine their rest-frame optical emission, particularly the diagnostic lines of \OIII{}~$\lambda 5007$ \AA{} and \OII{}~$\lambda3727$ \AA{}. Initial results from MOSFIRE were presented in \cite{nakajima2016} but those data have been enlarged in the present paper to take account of the associated imaging data taken with HST.

\subsection{Hubble Space Telescope Data}

The HST campaign (GO 14747, PI: Robertson) was conducted between UT 14th May 2017 - 20th December 2017 comprising four $F160W$ pointings with WFC3/IR of 1 orbit each (0.7 hrs) and three WFC3/UVIS $F336W$ pointings of 20 orbits each in five exposures per pointing (16 hrs, see Table \ref{table:HST_depths}). Together with the narrow band images taken with Subaru, this strategy allows us to compare prospective Lyman continuum leakage in the $F336W$ filter with associated signals in \Lya{} and the rest-frame optical continuum.

\begin{figure*}
\centerline{\includegraphics[width=\textwidth]{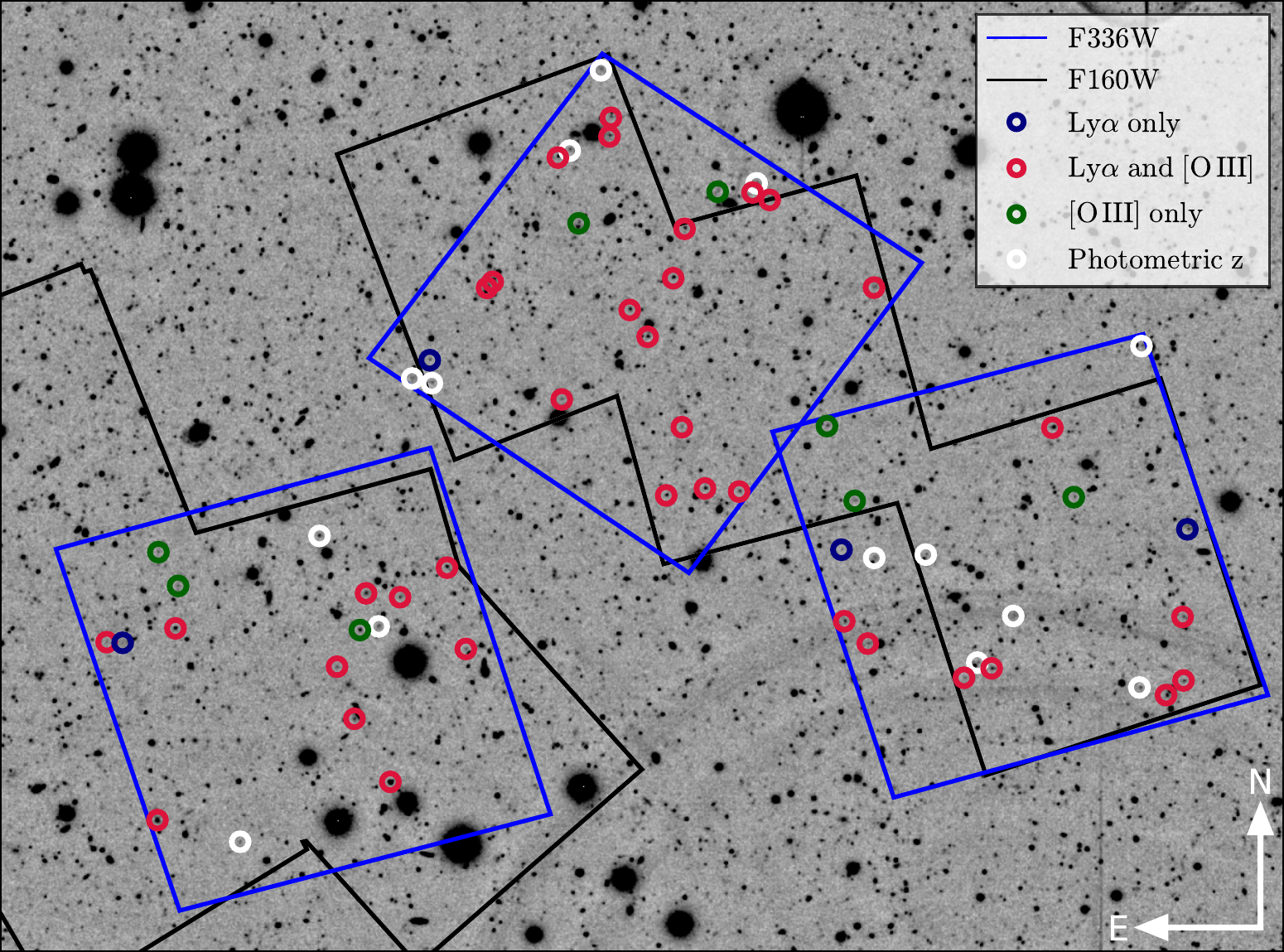}}
\caption{NB497 image showing the coverage of target Lyman alpha emitters and Lyman break galaxies from the full Subaru narrow-band selected sample within the three WFC3/UVIS $F336W$ pointings. The area covered by the smaller WFC3/IR $F160W$ pointings are delineated with overlapping black squares. The targets are color-coded according to their information content as follows: (red) \Lya{} spectroscopic redshift and \OIII{} line flux, (blue) \Lya{} spectroscopic redshift only, (green) \OIII{} spectroscopic redshift only, (white) photometric data only. For detailed statistics see Table \ref{table:summary}.}
\label{fig:map}
\end{figure*}

Within the coverage of the $F336W$ imaging, there are 61 sources from the original Subaru sample of which 54 are LAEs and 7 are LBGs. Although only 41 of the 54 LAEs have spectroscopic redshifts, we can exploit the remaining 13 narrow-band selected sources given contamination from foreground emitters has been shown to be negligible in practice \citep{matsuda2005,matsuda2006,yamada2012b,yamada2012}. A summary of the statistical sample is given in Table \ref{table:summary} and their distribution in the 3 WFC3 fields is shown in Figure \ref{fig:map}.

\
\subsubsection{WFC3/UVIS Reduction}

As described by \citet{rafelski2015a}, the standard HST pipelines may not result in science images appropriate
for the analysis of faint detections in WFC3/UVIS images. Correlated noise and other residual structure in the
images can lead to the misidentification of artefacts as low signal-to-noise sources, which must be
carefully addressed in any rigorous analysis.
For the Hubble Deep UV Legacy Survey,
\citet{oesch2018a} developed an approach to address the WFC3/UVIS data quality issue, described
in their Section 4.4 which we emulate here.
The goal is to construct a sky dark frame from the dark-corrected $F336W$ science exposures reduced using the STScI pipeline,
and then subtract the sky dark from each exposure to ameliorate the correlated noise and residual structure.
First, sources were identified in the original science images resulting from the STScI pipeline
by running Source Extractor \citep{bertin1996} with a detection
threshold of $1.5\sigma$ designed to capture genuine faint sources above typical noise fluctuations
in the images.
Based on the segmentation map of each pointing, all the pixels containing sources were
masked and replaced by simulated noise computed using the average and standard deviation of
background measured from nearby pixels around the sources.
A background frame was thereby extracted from each of the 15 science exposures and
then combined via median stacking to create the sky dark frame.
The sky dark was subtracted from each of the 15 exposures, and these sky dark-subtracted
frames were then combined into our final science images by creating median
stacks for each of the three pointings using {\tt SWarp} \citep{bertin2010swarp}.

As an indication of the importance of our implementation of the additional sky dark
subtraction, we compared our subsequent source identification and analysis with
respect to one originally made using the STScI pipeline products. In addition to
a significantly improved image quality, we find that our brightest $F336W$
sources (${\geq}4\sigma$ significance) are robust to the choice of reduction, but
the flux associated with these sources can vary substantially between reductions
(from decreasing by 50\% to increasing by 20\%). We further note that after the
original submission of this paper, updated versions of the STScI science products
became available (in January 2019) correcting an issue whereby incorrect calibration
products had been applied during the processing of WFC3/UVIS data for our program
GO 14747. The final sky darks and science images used herein have been corrected for this issue,
which also influenced the noise properties of the STScI products and the flux of
objects measured in uncorrected data. We mention these issues to highlight the
challenging systematics associated with analyzing WFC3/UVIS images at faint
flux levels, and to motivate our conservative sample definition where we restrict
our Lyman continuum candidate designation to ${\geq}4\sigma$ $F336W$ sources.

To accurately compare $F336W$ detections with signals in other bands, all HST images were astrometrically aligned with the most appropriate Subaru images using custom
software and verified with
visual inspection of hundreds of sources in each image. Sources were then extracted in the $F336W$, Subaru NB497 (\Lya{}) and $F160W$ images using a Python script based on the SEP tool\footnote{\url{https://github.com/kbarbary/sep/tree/v1.0.x}} \citep{bertin1996, barbary2016sep}.

 \begin{table}
 \centering
 \caption{HST imaging observations}
 \label{table:HST_depths}
 \renewcommand{\arraystretch}{1.4}
 \setlength{\tabcolsep}{0.45em}
 \begin{threeparttable}
 \begin{tabular}{@{}lcccccc@{}}
   \hline
   \hline
   Filter & Pointing & $\lambda_{\rm eff}$ & PSF\textsuperscript{a} & Pixel & Exposure & Depth\textsuperscript{b}\\
   & & & & Scale & & \\
   & & (\AA) &
   ($^{\prime\prime}$) &
   ($^{\prime\prime}$/pixel) &
   (s) & (AB)\\
   \hline
	$F336W$& 1 & 3355 & 0.081 & 0.040 & 57845 & 30.24 \\
    $F336W$& 2 & 3355 & 0.081 & 0.040 & 57845 & 30.32 \\
    $F336W$& 3 & 3355 & 0.081 & 0.040 & 57845 & 29.41 \\
    $F160W$ & All & 15369 & 0.151 & 0.128 & 2612 & 27.61\\
   \hline
 \end{tabular}
 \begin{tablenotes}
 \small
 \item \textsuperscript{a} The FWHM of the PSF.
 \item \textsuperscript{b} The $3\sigma$ limiting magnitude using an aperture with a diameter 1.5 times the size of the PSF FWHM (e.g. $0.12^{\prime\prime}$ for the $F336W$ images).
 \end{tablenotes}
 \end{threeparttable}
\end{table}

\subsection{Additional Photometry}

All the LACES targets are covered with the plentiful, deep multi-wavelength photometric data in the SSA22 field. We utilize the photometric data including Subaru, CFHT, UKIRT and Spitzer/IRAC imaging data in addition to the HST/$F160W$ photometry to constrain the nature of the stellar populations of the LACES objects via an SED fitting analysis (Section \ref{sec:SEDs}). Table \ref{table:opt_nir_photometry} gives the details of the additional photometric data.

We perform photometry of the bands listed in Table \ref{table:opt_nir_photometry} on the LACES sources using TPHOT (v2.0; \citealt{merlin2015,merlin2016}). Briefly, we use the HST/$F160W$ image as a high resolution reference image and extract the spatial and morphological information of objects within a radius of $\sim 25^{\prime\prime}$ from each of the LACES sources. Using this information and a kernel carefully created to convolve the high resolution image to have the PSF of the lower resolution ground-based images, TPHOT produces templates of the objects in the low resolution image. TPHOT then varies the brightness of each of the templates to match the global observed flux in the low resolution image. In this way we can accurately measure total fluxes from the low resolution images in Table \ref{table:opt_nir_photometry} by removing light from nearby contaminating sources. For sources not detected or not covered by the HST/$F160W$ image, we adopt aperture photometry with a $2^{\prime\prime}$ diameter aperture for the Subaru, CFHT and UKIRT images and $3^{\prime\prime}$ for the IRAC data and fix the position of the aperture determined using the NB497 detection. The aperture magnitudes are then converted into total magnitudes using aperture correction values, which are estimated from differences between aperture and total magnitudes for point sources. We have confirmed that the two methods return a consistent SED within the $1\sigma$ uncertainties for isolated objects.


 \begin{table}
 \centering
 \caption{Summary of optical and NIR imaging data}
 \label{table:opt_nir_photometry}
 \renewcommand{\arraystretch}{1.4}
 \setlength{\tabcolsep}{0.45em}
 \begin{threeparttable}
 \begin{tabular}{@{}lcccccc@{}}
   \hline
   \hline
   Filter &
   Observatory &
   PSF &
   Depth &
   Reference \\
     &
     &
    ($^{\prime\prime}$)\textsuperscript{a} &
    (mag)\textsuperscript{b} &
    \textsuperscript{c} \\
   \hline
   $u^{\star}$ &
    CFHT &
     $1.0$ &
     $26.0$ &
     (1) \\
   $B$ &
    Subaru &
    $1.0$ &
    $26.5$ &
    (2), (3), (4) \\
   NB$497$ &
    Subaru &
    $1.0$ &
    $26.2$ &
    (2), (3), (4) \\
   $V$ &
    Subaru &
    $1.0$ &
    $26.6$ &
    (2), (3), (4) \\
   $R$ &
    Subaru &
    $1.1$ &
    $26.7$ &
    (2), (4) \\
   $i^{\prime}$ &
    Subaru &
    $1.0$ &
    $26.4$ &
    (2) \\
   $z^{\prime}$ &
    Subaru &
    $1.0$ &
    $25.7$ &
    (2) \\
   $J$ &
    UKIRT &
    $0.9$ &
    $23.5$ &
    (5) \\
   $K$ &
    UKIRT &
    $0.8$ &
    $23.1$ &
    (5) \\
   $[3.6]$ &
    Spitzer &
    $2.0$ &
    $22.2$--$24.7$ &
    (6) \\
   $[4.5]$ &
    Spitzer &
    $2.0$ &
    $22.2$--$24.4$ &
    (6) \\
   \hline
 \end{tabular}
 \begin{tablenotes}
 \small
 \item \textsuperscript{a} The FWHM of the PSF.
 \item \textsuperscript{b} The $5\sigma$ limiting magnitude using an aperture with a diameter of $3^{\prime\prime}$ for the IRAC $3.6$ and $4.5\,\mu$m bands and $2^{\prime\prime}$ for the other bands.
 \item \textsuperscript{c} (1) \citealt{hayashino2019a}; (2) \citealt{hayashino2004}; (3) \citealt{yamada2012}; (4) \citealt{matsuda2005}; (5) \url{http://wsa.roe.ac.uk/}; (6) \url{http://sha.ipac.caltech.edu/applications/Spitzer/SHA/}
 \end{tablenotes}
 \end{threeparttable}
\end{table}



\subsection{Near-Infrared Spectroscopy}
\label{sec:NIR_spectroscopy}

In addition to the initial 2015 campaign reported in \cite{nakajima2016}
which targeted only one MOSFIRE pointing (referred to here as mask 1) in SSA22, we have now completed spectroscopy of three further pointings (masks 2-4) within the HST covered area (Figure \ref{fig:map}). The new observations were taken on UT July 31, August 1 and October 10 2017 in photometric conditions with seeing ranging from 0.5-0.9 arcsec in the summer months to 0.3-0.5 arcsec on the more recent run. Spectra were obtained in both the K band (sampling \OIII{} and \Hb{} at a spectral resolution $R\simeq 3600$) and H band (sampling \OII{} at $R\simeq 3700$) for the masks 1 and 2, while only in the K band for the masks 3 and 4. Individual exposures of 180 sec (120 sec) were taken in K (H) with a AB nod sequence of 3.0 arcsec separation. The total on-source exposure times ranged from 2 to 3 hours with some sources included on both mask 2 and mask 4.

Data reduction was performed using the MOSFIRE DRP\footnote{https://keck-datareductionpipelines.github.io/MosfireDRP/} in the manner described in \cite{nakajima2016}. Briefly, the processing includes flat fielding, wavelength calibration, background subtraction and combining the nod positions. Wavelength calibration in H was performed using OH sky lines and in K a combination of OH lines and Neon arcs were used. Flux calibration and telluric absorption corrections were obtained from A0V Hipparcos stars observed at similar air masses as well as via relatively bright stars ($K_{\mathrm{Vega}}$ = 15.5--16.5) included on each of mask.

We measured the \OII{} and \OIII{} line fluxes by fitting a Gaussian profile to each line using the IRAF task {\tt spec fit}. In deriving the rest-frame equivalent width (EW) of \OIII{}, we used the measured $F160W$ in conjunction with a mean spectral energy distribution of $z\simeq 3.1$ LAEs \citep{ono2010} to determine the continuum flux in the vicinity of the line near $2.05 ~\rm \mu m$. We investigated the effect of dust corrections using individually derived $E(B-V)$ values for each object (see Section \ref{sec:analysis}) but as our LAEs are mostly dust-free the corrections were small. Table \ref{table:summary} summarizes the statistics of the \OII{} and \OIII{} detections. The full catalog of line fluxes and equivalent widths will be reported later as the spectroscopic campaign continues.

In total, 51 of the 61 sources in the 3 WFC3 fields have \OIII{} detections or upper limits. The coverage of \OII{} is less complete at present, with roughly half of the \OIII{} sample containing \OII{} data or upper limits (see Table \ref{table:summary}).

\begin{table}
 \centering
 \caption{Summary of the LACES sample}
 \label{table:summary}
 \renewcommand{\arraystretch}{1.4}
 \begin{tabular}{@{}lccc@{}}
   \hline
   \hline
   $N^o$ of objects &
   LAEs &
   LBGs &
   Total\\
   \hline
   Within the HST area & $54$ & $7$ & $61$\\
   Spectroscopic redshift & $41$ & $6$ & $47$ \\
   $F160W$ coverage or limits & $45$ & $7$ & $52$\\
   \,[O\,{\sc iii}] or limits & $46$ & $7$ & $53$\\
   \,[O\,{\sc ii}] or limits (with [O\,{\sc iii}] identified) &
$23$ & $4$ & $27$\\
   $F160W$ coverage \& both [O\,{\sc iii}]+[O\,{\sc ii}] data &
$23$ & $4$ & $27$\\
   \hline
 \end{tabular}
\end{table}

\medskip


\section{Lyman Continuum Candidates}
\label{sec:candidates}

We now discuss the procedure adopted to decide which SSA22 sources show promising evidence of Lyman continuum leakage in the HST $F336W$ filter. The key issues include the optimum aperture for measuring the $F336W$ flux, the photometric significance of any detection, the spatial coincidence with signals in other bands and the possibility of foreground contamination. We also discuss the nature of those sources where no significant $F336W$ flux is seen and examine the possibility of providing a statistical detection on the basis of a stacking analysis.

\subsection{Detections}
\label{sec:detections}


We first constructed a mosaic of all 61 targets with HST $F336W$ coverage, comparing the location and morphology of possible $F336W$ detections with images in Subaru NB497 (\Lya), $R$ and HST $F160W$. Five authors (MT, BER, TF, RSE, DPS) examined this mosaic for potential $F336W$ detections. Although the Subaru \Lya{} image offers a natural astrometric reference point, as a ground-based image with $1$ arcsecond seeing it is less useful than the HST $F160W$ image which samples the rest-frame optical light and can reveal complex source morphology. In practice we found it helpful to overlay a $F160W$ contour over the $F336W$ image to evaluate spatial coincidence.

The photometric significance of possible $F336W$ detections was also taken into account on the assumption that LyC signals would be mostly unresolved with HST. Fluxes were measured in an aperture whose diameter is 1.5 times the $F336W$ point spread function (i.e. 3 WFC3/UVIS pixels, 0.12 arcsec, 0.91 kpc at $z=3.1$). This aperture is more sensitive in discovering candidates than adopting a (larger) matched aperture across all the photometric bands that would introduce unnecessary noise. For the few sources that show extended emission in the $F336W$ image we only measure a signal from the brightest peak, possibly underestimating the true $F336W$ flux and \fesc{}.

To evaluate the completeness of our search and provide useful upper limits for the non-detections, we masked all the detected sources above a threshold of 1.5$\sigma$ of the background noise. Fake sources with a Gaussian profile corresponding to the PSF of the images and known magnitudes were inserted into the unmasked regions and the detection algorithm re-run. In this manner we determined a 75\% completeness limit of $F336W \rm (AB) = 29.90$. We verified this noise limit with that determined from aperture measures in the vicinity of each target.

For each LyC candidate the noise level was measured locally in a $4 \times 4$ arcsecond region around the target. The targets, neighboring objects and any signal $5\sigma$ above the noise level were masked in the postage stamps. Using these segmentation maps, apertures 1.5 times the PSF, the same size used to measure the flux, were randomly distributed and the $1\sigma$ noise level calculated from the random placement of apertures on the image.


\subsection{Gold and Silver Subsamples}\label{sec:gold_silver}

\begin{figure*}
\begin{center}
\includegraphics[width=5in]{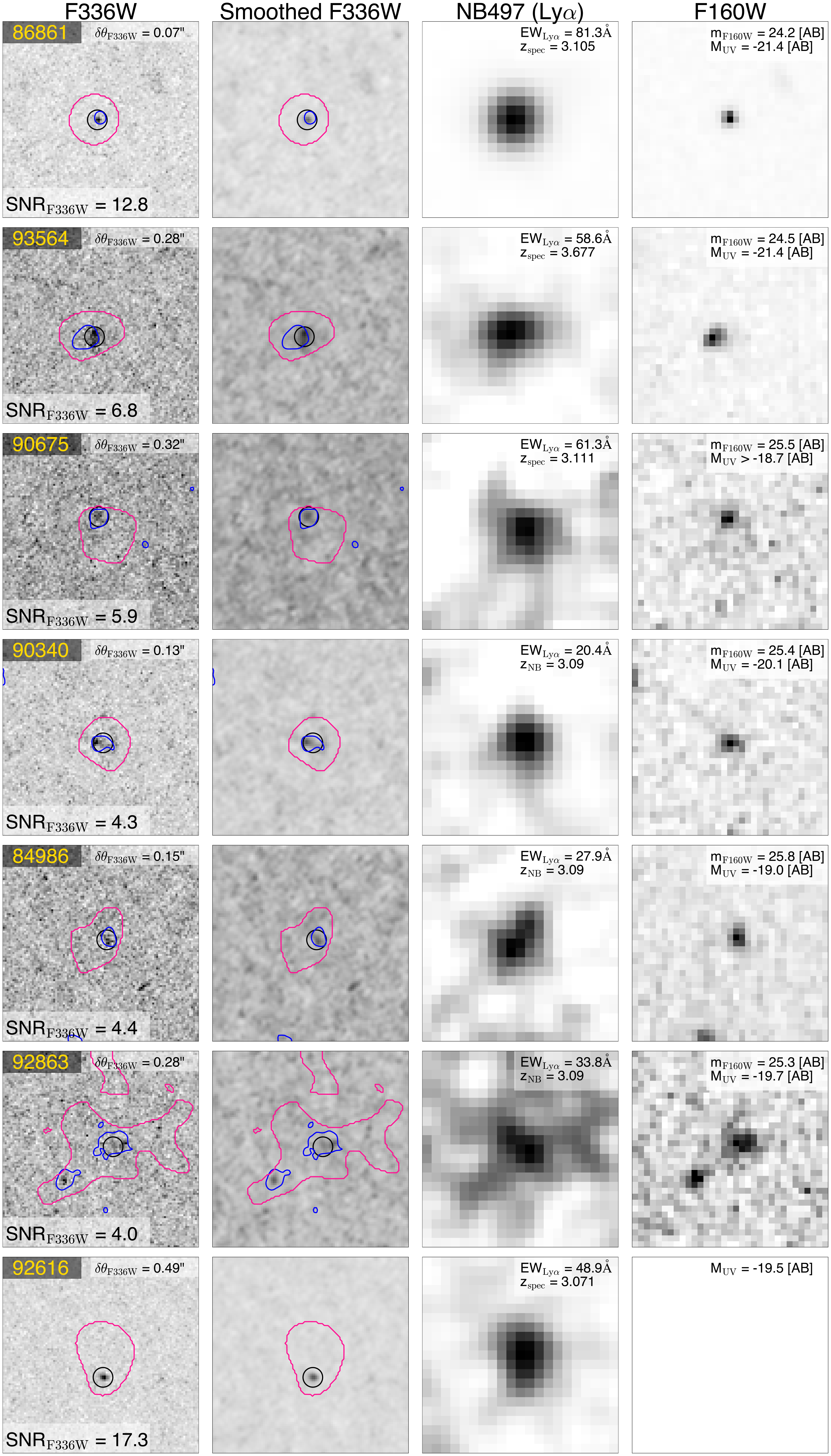}
\end{center}
\caption{Mosaic of astrometrically aligned 4 $\times$ 4 arcsec images for sources with high signal to noise $F336W$ detections (see text for discussion of the detection procedure) comprising the Gold subsample. From left to right each panel displays (i) the background-subtracted $F336W$ image overlaid with contours from the narrow band 497nm (magenta) and $F160W$ (blue) images and the location of the corresponding $F336W$ source (black circle; 0.02'' radius), (ii) the former smoothed, (iii) the Subaru narrow band 497 nm image, and (iv) the $F160W$ image (where available), along with a summary of physical properties where the SNR refers to the F336W detection. All of our objects are LAEs except for 86861 which is an LAE-AGN.}
\label{fig:gold_mosaic}
\end{figure*}

\begin{figure*}
\begin{center}
\includegraphics[width=5in]{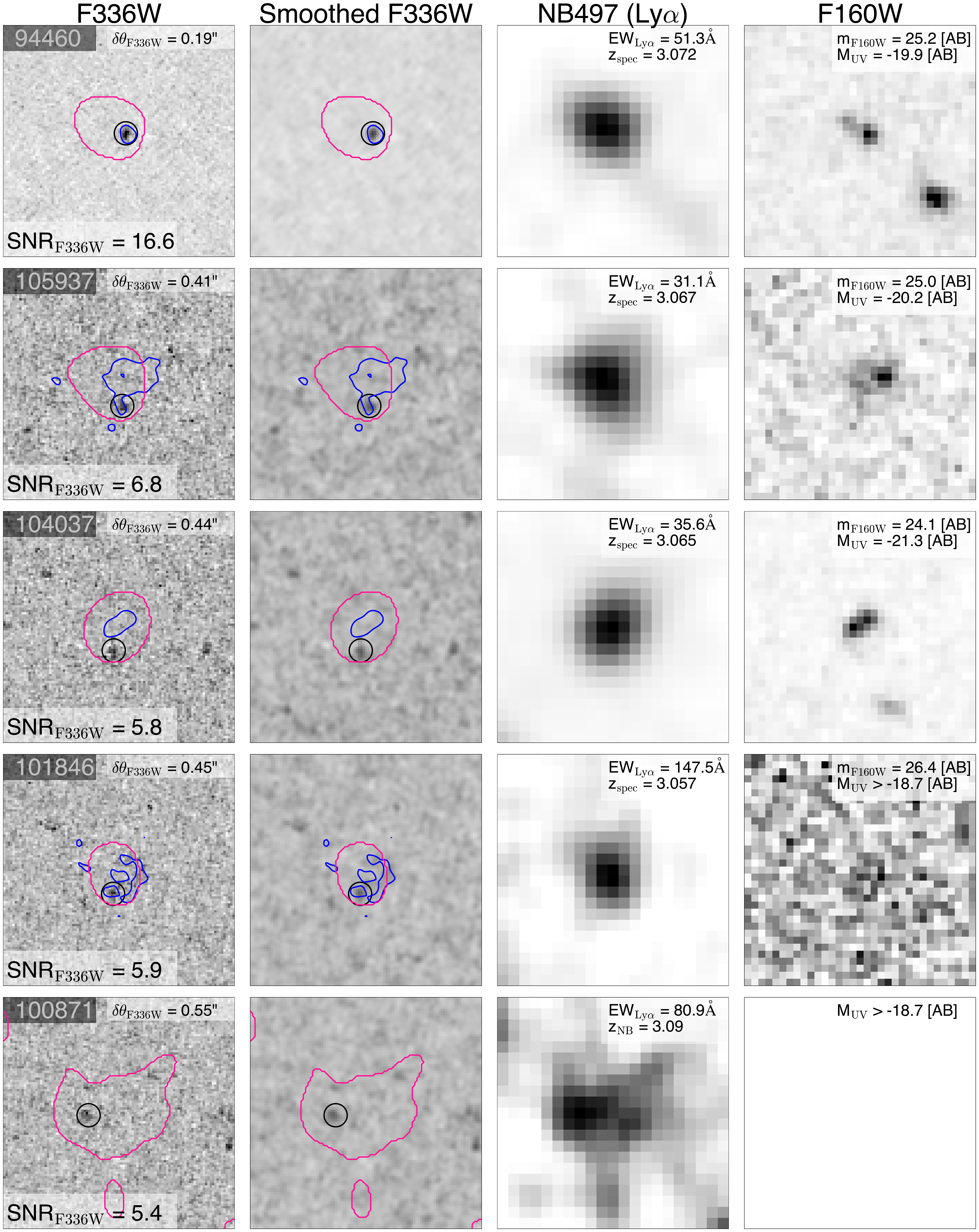}
\end{center}
\caption{Same as Figure \ref{fig:gold_mosaic}, but for the Silver subsample.}
\label{fig:silver_mosaic}
\end{figure*}

We have conservatively divided our detections into Gold and Silver subsamples in order to distinguish between cases where we are respectively convinced and reasonably sure the detected LyC flux is associated with the target galaxy. In our subsequent analysis it will be helpful to examine trends separately between the Gold and Silver subsamples as well as with those for the non-detections.
We show in Figures \ref{fig:gold_mosaic} and \ref{fig:silver_mosaic} postage stamps of $4 \times 4$ arcsec for our Gold and Silver subsamples, defined according to the criteria below. Together they comprise 12 sources for which reasonably convincing $F336W$ detections were found by the procedure outlined in Section \ref{sec:detections}. To assist in recognizing the detections, we also show the $F336W$ images smoothed with a 2D Gaussian with $1\sigma$ equal to 1 pixel. We also show an overlay of the $F160W$ and \Lya{} narrow band contours on the $F336W$ images to illustrate the spatial coincidence of the optical continuum and LyC flux.
Table \ref{tbl:Gold} summarizes the Gold and Silver subsamples alongside their photometric and spectroscopic properties.

To qualify for the Gold sample, targets must satisfy three criteria:

\begin{enumerate}

\item Availability of a spectroscopic redshift for the target or no evidence that the target may be an interloper. A redshift may seem an essential requirement but the probability a NB497 excess which leads to a selected $z=3.1$ LAE is contaminated by a foreground emission line is very low
\citep{matsuda2005,matsuda2006,yamada2012b,yamada2012} so only if there is some spectroscopic evidence for an interloper would the candidate be rejected.

\item The $F336W$ flux must be spatially coincident (to within 0.6 arcsec) with the core of the $F160W$ flux or the \Lya{} centroid. In cases where the $F160W$ image reveals substructure, there is a danger the $F336W$ detection is coincident with an interloper. Although we will show this possible contamination is unlikely, such a configuration merits demotion to the Silver subsample.

\item The $F336W$ detection has a signal to noise ${\geq}4$  as evaluated by the process discussed in the next subsection.

\end{enumerate}

Using these criteria we select 7 Gold candidates shown in Figure \ref{fig:gold_mosaic}. Target IDs 86861, 93564, 90675 and 92616 are all spectroscopically confirmed at $z\gtrsim 3.07$ and are coincident with compact $F160W$ regions, except for 92616 where there is no $F160W$ imaging. In the latter case, the $F336W$ centroid is coincident with the ground-based {\it Subaru} optical broad band counterpart and inside the more extended \Lya{} emission.
IDs 84986 and 90340 are considered worthy of inclusion because in both the MOSFIRE spectrum no other lines were detected suggesting an interloper is unlikely. IDs 92863 and 100871 were not targeted with IR spectroscopy.

\begin{table*}
\centering
\caption{Properties of the LyC Leaking Candidates}
\label{tbl:Gold}
\renewcommand{\arraystretch}{1.25}
\begin{tabular}{@{}lccccccccccc@{}}
\hline
ID
& $f_{\rm Lyc}$
& SNR
& $M_{\rm UV}$
& EW(Ly$\alpha$)
& $z_{{\rm sys}}$
& $\Delta v_{{\rm Ly}\alpha}$
& EW([O\,{\sc iii}])
& [O\,{\sc iii}]$/$H$\beta$
& R23
& [O\,{\sc iii}]$/$[O\,{\sc ii}]
\\
& ($10^{-9} \rm Jy$)
&
&
& (\AA)
&
& (km\,s$^{-1}$)
& (\AA)
&
&
&
\\
\hline
\multicolumn{11}{c}{Gold Sample}\\
\hline
$\rm 86861^{*}$
& $13.6\pm 1.1$
& $12.8$
& $-21.36\pm 0.03$
& $81^{+2}_{-2}$
& $3.1054$
& $313.3$
& $295.4\pm 13.3$
& $9.7\pm 0.7$
& $9.7\pm 0.7$
& ${>}6.2$
\\
93564
& $6.2\pm 0.9$
& $6.9$
& $-21.35\pm 0.05$
& $58^{+6}_{-6}$
& $3.6770$
& $574.3$
& $1040.1\pm 33.7$
& $8.6\pm 0.8$
& $9.5\pm 0.9$
& $10.1\pm 0.9$
\\
90675
& $4.6\pm 0.8$
& $5.3$
& ${>}{-}18.72$
& ${>}61$
& $3.1110$
& $-3.6$
& ${<}66.3$
& ${<}0.4$
& --
& --
\\
90340
& $4.3\pm 1.0$
& $4.3$
& $-20.08\pm 0.11$
& $20^{+7}_{-6}$
& --
& --
& ${<}55.0$
& --
& --
& --
\\
84986
& $4.8\pm 1.1$
& $4.4$
& $-18.96\pm 0.33$
& $27^{+13}_{-10}$
& --
& --
& ${<}210.1$
& --
& --
& --
\\
92863
& $4.1\pm 1.0$
& $4.0$
& $-19.67\pm 0.16$
& $33^{+12}_{-9}$
& --
& --
& --
& --
& --
& --
\\
92616
& $14.5\pm 0.8$
& $17.3$
& $-19.51\pm 0.19$
& $48^{+12}_{-10}$
& $3.0714$
& $253.3$
& --
& ${>}6.2$
& --
& --
\\
\hline
\multicolumn{11}{c}{Silver Sample}\\
\hline
94460
& $16.9\pm 1.0$
& $16.6$
& $-19.88\pm 0.13$
& $51^{+8}_{-7}$
& $3.0723$
& $157.5$
& $384.9\pm 22.6$
& $8.4\pm 1.8$
& $8.4\pm 1.8$
& ${>}10.9$
\\
105937
& $6.2\pm 0.9$
& $6.8$
& $-20.22\pm 0.09$
& $31^{+7}_{-6}$
& $3.0668$
& $143.7$
& $103.4\pm 15.7$
& $2.4\pm 0.5$
& --
& --
\\
104037
& $5.2\pm 0.9$
& $5.8$
& $-21.34\pm 0.03$
& $35^{+2}_{-2}$
& $3.0650$
& $166.7$
& $791.7\pm 17.4$
& $8.2\pm 0.3$
& $9.6\pm 0.4$
& $5.9\pm 0.2$
\\
101846
& $5.0\pm 0.9$
& $5.9$
& ${>}{-}18.67$
& ${>}147$
& $3.0565$
& $232.1$
& ${>}165.6$
& ${>}1.5$
& --
& --
\\
100871
& $4.8\pm 0.9$
& $5.4$
& ${>}{-}18.71$
& ${>}80$
& --
& --
& --
& --
& --
& --
\\
\hline
\end{tabular}

\\
\vspace{-1mm}
\begin{flushleft}
\scriptsize
$\rm SNR\geq 4$ is required for a detection. The columns (numbered) denote the following:
(1) LyC flux.
(2) LyC signal-to-noise ratio.
(3) Absolute UV magnitude.
(4) Rest equivalent width (EW) of Ly$\alpha$. For the $z\simeq 3.1$ objects, the EW is estimated from the BV$-$NB497 color in conjunction with the Ly$\alpha$ redshift. The EW of LAE93564 is derived from spectroscopy.
(5) Systemic redshift measured from the [O\,{\sc iii}] and H$\beta$ line(s).
(6) Velocity offset of Ly$\alpha$, $(z_{{\rm Ly}\alpha}-z_{\rm sys})/(1+z_{\rm sys})\times c$.
(7) Rest EW of [O\,{\sc iii}] $\lambda\lambda 5007, 4959$. The associated continuum is estimated from HST/$F160W$ photometry, which is translated into the flux density at $5000$\,\AA\ with the typical SED of $z \sim 3$ LAEs \citep{ono2010}.
(8) [O\,{\sc iii}] $\lambda\lambda 5007, 4959$$/$H$\beta$.
(9) R23-index.
(10) [O\,{\sc iii}] $\lambda\lambda 5007, 4959$$/$[O\,{\sc ii}] $\lambda 3727$.
No reddening correction has been applied to the Oxygen and ${\rm H}\beta$ values presented here as
the reddening correction is very small (see Table \ref{table:sed_params}). The
$*$ denotes the one LAE-AGN in our sample. Objects 86861, 93564 and 94460 are reported respectively as AGN04, LBG01 and LAE06 in \cite{micheva2017,micheva2015}.
\end{flushleft}
\end{table*}

We show the Silver subsample in Figure \ref{fig:silver_mosaic}. These Silver sources (all
${\geq}4\sigma$)
have more complicated morphologies or less conclusive spectroscopic information, but otherwise would qualify
for the Gold subsample. ID 94460 is placed in the Silver sample because, in addition to \Lya{}, \OIII{} and $H\beta$ all at $z=3.07$, a spatially-offset emission line inconsistent with this redshift
was found which may indicate a contaminating source. IDs 104037 and 105937 fall into the Silver sample due to their extended $F160W$ regions.
Although 100871 also has no $F160W$ imaging and 101846 has only a faint $F160W$ source, their $F336W$ signals are coincident with the peak of \Lya{} emission.


\subsection{Spatial Offsets}\label{sec:spatial_offsets}

In considering the validity of our sample we now examine two further criteria. The first is the distribution of separations between the $F336W$ centroid and that for $F160W$ and Subaru \Lya{}. Although a precise spatial coincidence of LyC leakage and UV/optical continuum and/or \Lya{} emission is desirable, previous studies have already found the LyC emission is occasionally offset from that of \Lya{}~\citep{iwata2009,mostardi2013,mostardi2015,micheva2015} with a median separation of $\sim 5$ proper kpc reported in \cite{mostardi2015}.

\begin{figure}
\centering
\includegraphics[width=\linewidth]{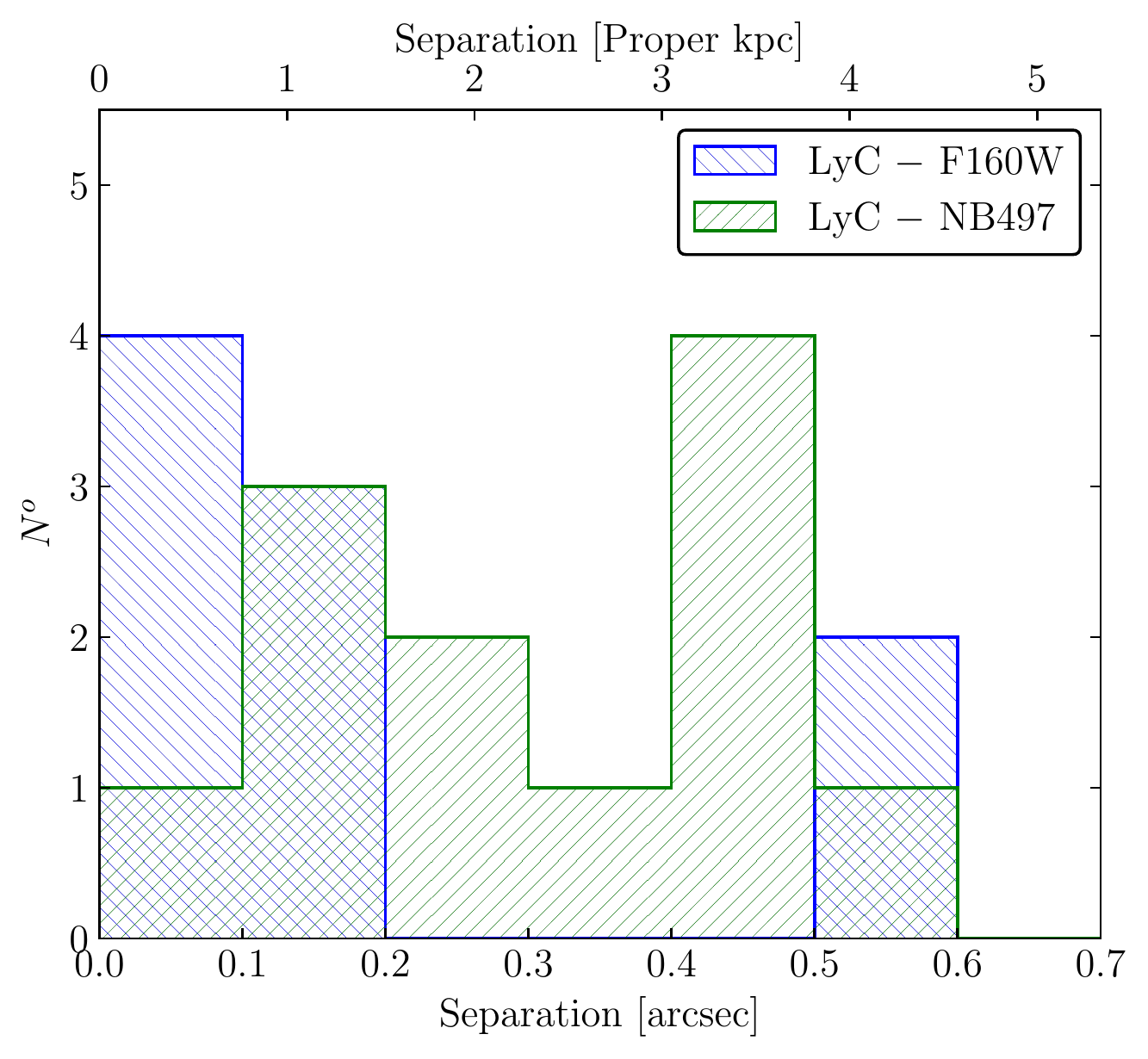}
\caption{Distribution of separations between the peak of the LyC emission and the peak of both the \Lya{} (12 cases) and $F160W$ emission (when available, 9 cases).}
\label{fig:seps}
\end{figure}

Figures \ref{fig:gold_mosaic} and \ref{fig:silver_mosaic} indicate that, for the majority of our targets, the $F336W$ centroid falls perfectly within the contours from the rest-frame optical ($F160W$) continuum where available. Figure \ref{fig:seps} shows the distribution of spatial offsets between $F336W$ and both the \Lya{} and $F160W$ centroids. For 75\% of the candidates, the separation between the LyC and $F160W$ centroid is less than 0.4 arcseconds (3 proper kpc at $z=3.1$). For reference $1^{\prime\prime}$ at $z=3.1$ corresponds to $7.6$ proper kpc. The angular resolution of the ground-based NB497 images is naturally worse. However for all of our candidates the $F336W$ emission lies within $0.6^{\prime\prime}$ (4.6 proper kpc) of the \Lya{} centroid which we consider satisfactory given the seeing in the Subaru image is $\simeq$1.0 arcsec.

In fact, targets with larger LyC - $F160W$ separations tend to have extended $F160W$ or \Lya{} emission. In these cases the LyC emission is still coincident but, due to the extended nature of the source, it can fall further from the centroid in the NB497 or $F160W$ bands. If LyC photons are emitted from regions occupied by young stars, then LyC may reasonably
lie closer to the rest-frame optical compared
with \Lya{} that may be resonantly scattered. These small separations are encouraging and lead us to believe the putative $F336W$ detections are due to LyC photons emitted from LAEs at $z\simeq 3.1$.

\subsection{Foreground Contamination}\label{sec:contamination}

Although we have attempted to isolate candidates whose $F336W$ detections may arise from foreground contaminants, we can estimate statistically the likelihood of interlopers from luminosity functions of lower redshift galaxies.
The relevant calculation requires, as input, the aperture within which LyC flux is searched. As a result, estimates of contamination must account for the possibility
that the most active star-formating regions from which LyC photons are emitted could be spatially offset from the bulk of the stars and gas in the galaxy.
This effect has been discussed in both ground-based studies \citep{iwata2009, inoue2011, nestor2011,nestor2013} and those using HST \citep{mostardi2015}. Allowing for spatial offsets increases the effective aperture and hence increases the likelihood of foreground contamination. In addition, the likelihood of contamination decreases with the depth and resolution of the imaging data. Fortunately, in our case, deep WFC3 UVIS/$F336W$ imaging provides the best angular resolution possible and probes to depths of 30.2 AB magnitudes.

\begin{figure}
\centering
\includegraphics[width=\linewidth]{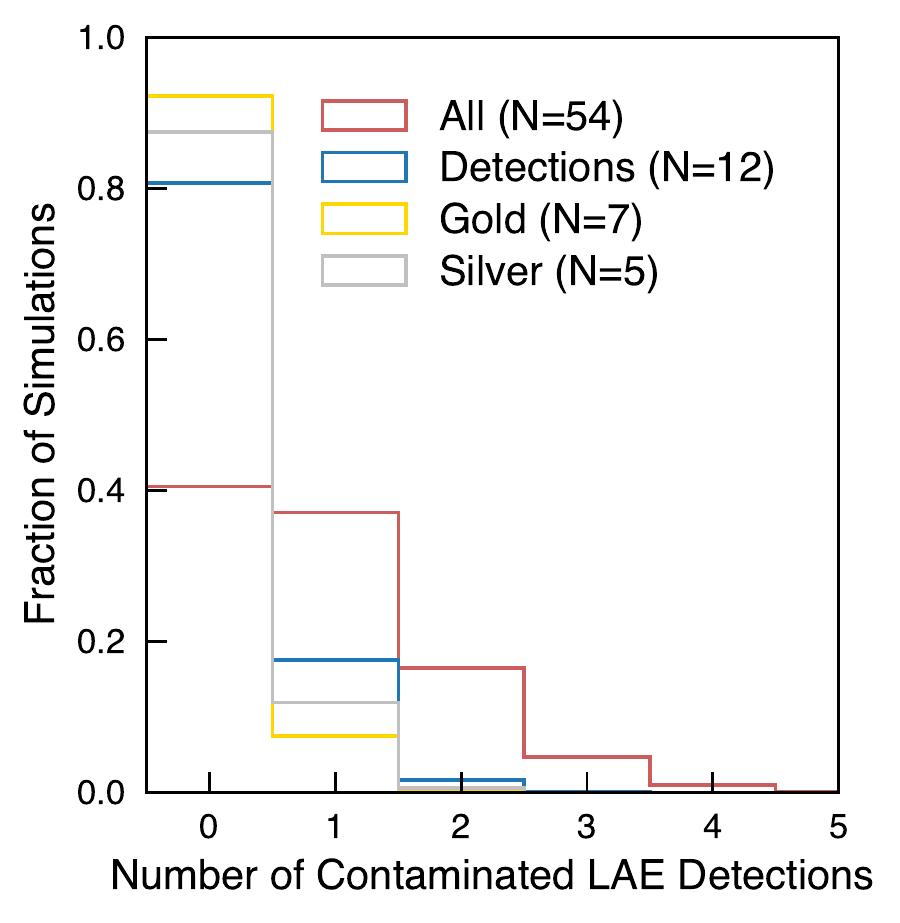}
\caption{Probability distribution of contamination in our sample of 12 detected objects (blue), and our Gold and Silver subsamples. In $98\%$ of cases $\leq 2$ of the 12 LyC detections are predicted to be affected by contamination. For the Gold subsample, we expect no contamination in $92\%$ of cases. For the Silver subsample, we expect no contamination in $87.5\%$ of cases. We also plot the number of contaminates expected in a sample of 54 LAEs, assuming the same LyC offset distribution as our detected sources (red). Even in this much larger sample, $99\%$ of cases have three or fewer possible contaminates.}
\label{fig:contamination}
\end{figure}

Following the method of \cite{vanzella2010}, we now calculate the probability a single source is contaminated and the probability $N$ of our 12 ${\geq}4\sigma$ detections suffer from foreground contamination.
We use the number counts per $\rm deg^{2}$ in \cite{vanzella2010} derived from the ultra-deep VIMOS U band imaging taken in the GOODS-S field \citep{nonino2009}. As our $F336W$ measurements are very deep we use the $3\sigma$ upper limits for the faintest U band magnitudes extrapolated to 30.5 AB \citep{vanzella2010}.
We adopt the distribution of offsets shown in Figure \ref{fig:seps}, for all our ${\geq}4\sigma$ detections, our Gold subsample, and our Silver subsample as apertures for considering possible
contamination from the foreground U band sources.
In Figure \ref{fig:contamination} we combine these individual estimates for contamination and run Monte Carlo simulations to show the probability that $N$ of the candidates could be contaminated. The probability that 0, 1, 2 or 3 of our 12 candidates could be contaminated is 80\%, 18\%, 1.6\% and ${<}0.01$\% respectively. Indeed, in 98\% of cases we estimate that $\leq 2$ of our LyC detections could be contaminated. Considering the Gold and Silver samples separately, we expect that these samples would
be pure in 92\% and 87.5\% of cases, respectively. We cannot rigorously perform the same analysis on the non-detections as we cannot measure the possible offsets between LyC and \Lya{} and $F160W$ centroids. However, if we assumed a similar distribution of offsets, applying the same analysis to all 54 LAEs would still predict far fewer potential contaminants than the number of detections we report for the LACES sample.
Quantitatively, we would expect that out of a possible 54 LAEs we would expect 0, 1, 2, 3, or 4 contaminates in 40.5\%, 37.1\%, 16.5\%, 4.72\%, or 0.98\% of cases. We can also use the observed distribution of candidates to estimate the maximum offset of contaminants. One can show that the maximum expected offset of contaminants in the \citet{vanzella2010} model is related to their mean offset by a geometrical factor of $\sim1.5$. If we adopt a maximum offset equal to $1.5\times$ the mean offset of the Gold sample, we would expect 0, 1, 2, 3, or 4 contaminates in 33.8\%, 37.0\%, 19.9\%, 6.99\%, or 1.8\% of cases assuming 54 possible LAEs.
We therefore have identified a robust sample of galaxies displaying Lyman continuum emission.

\subsection{Non-Detections}
\label{sec:non-detections}

The majority (42 of 54) of our LAEs have no clear $F336W$ detections above a signal to noise ratio of 4. To determine if these non-detections simply represent a tail of fainter signals, we can stack the non-detections to derive a statistical estimate of their mean $F336W$ flux. In this case, we must first consider how to register the
images, recognizing that the LyC flux may not always precisely coincide with either the $F160W$ or \Lya{} signals (Figure \ref{fig:seps}). We can evaluate the impact of such spatial offsets by conducting the same stacking experiment on those sources for which we see individual detections. By comparing the stack for the Gold and Silver subsamples based on different registrations ($F160W$ and \Lya{}), we can compare the loss in stacked signal compared to a direct sum of the registered $F336W$ detections.

For the following stacking procedure, we used custom software based on {\tt AstroPy} to perform a median stack centered on the position of either the $F336W$ peak (for the Gold and Silver subsamples), the Subaru NB497 \Lya{} peak and the HST $F160W$ peak. The $F336W$ frames were smoothed with a 1-pixel RMS width Gaussian before stacking to account for small relative offsets or astrometric uncertainties.

\begin{figure}
\centering
\includegraphics[width=\linewidth]{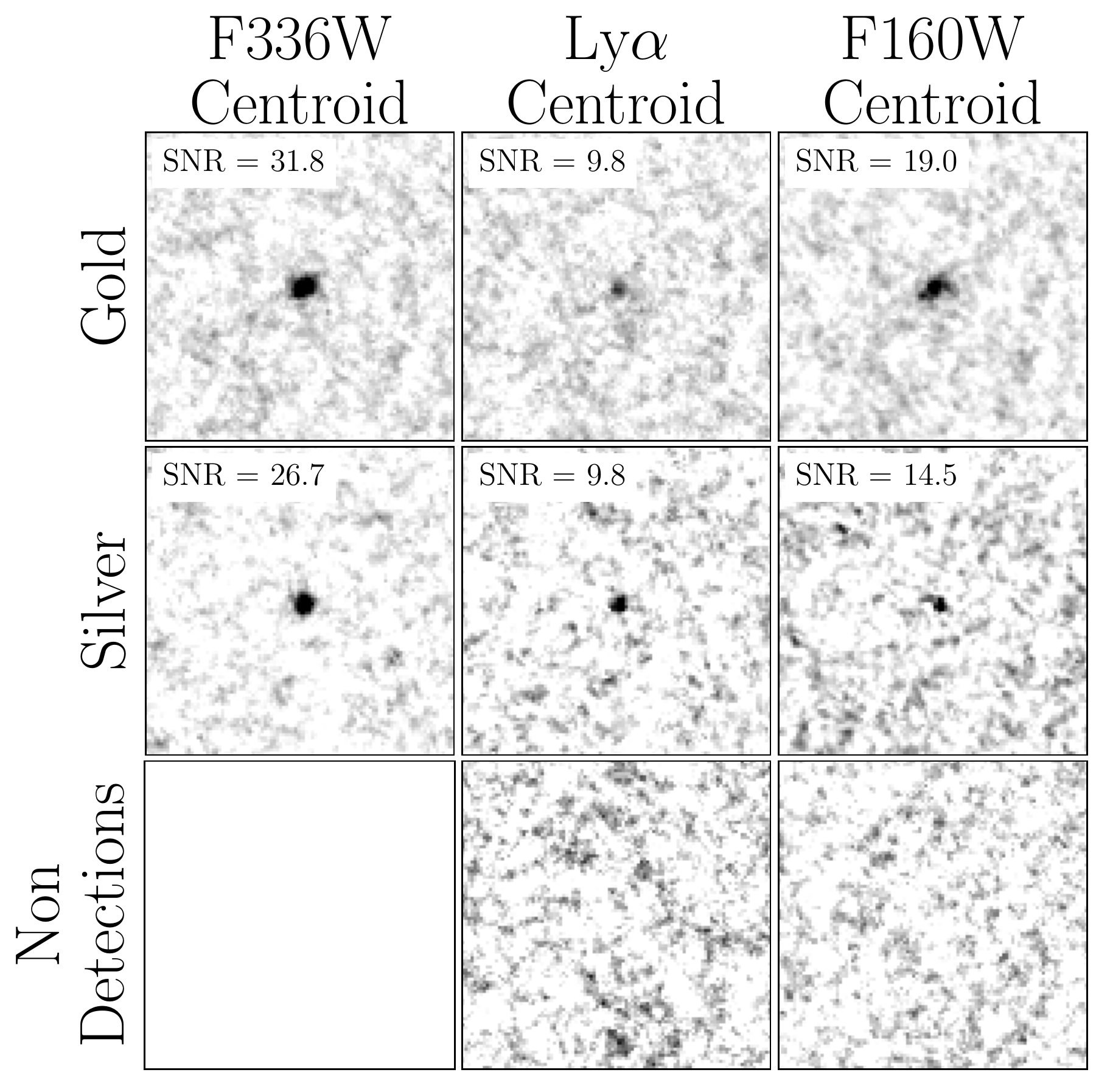}
\caption{Mosaic of $4\times 4$ arcsec images showing stacked $F336W$ images for the Gold (top row), Silver (middle row) and non-detections (bottom row) samples using different methods to center on the candidate galaxies. From left to right each panel displays the detections stacked using (i) $F336W$ centroid, (ii) \Lya{} centroid, (iii) $F160W$ centroid or \Lya{} centroid for cases where $F160W$ is unavailable.}
\label{fig:Det_stacks}
\end{figure}

The results for the detected (Gold and Silver) subsamples are shown in Figure \ref{fig:Det_stacks}. As expected using the $F160W$ centroid generally gives a better S/N ratio in the final stack and there is little degradation in signal compared to a direct summation of the $F336W$ signals, especially for the Gold subsample. This correspondence simply reflects the small spatial offsets involved. The signal-to-noise ratio of the Gold (Silver) $F336W$ stacks are $SNR=31.8$ (26.7) when using the $F336W$ centroid, $SNR=9.8$ (9.8) when using the \Lya{} centroid, and $SNR=19.0$ (14.5) when using the $F160W$ centroid. Strong stacked detections remain even without smoothing, with the Gold sample stacks showing $SNR=24.0$, 6.9, and 13.8 for when centroiding on $F336W$, \Lya{}, and $F160W$, respectively, and the Silver sample stacks showing similar significance.

Applying the same procedure now to $z\simeq 3.1$ LAEs not detected individually in $F336W$, we can only register using the \Lya{}  ($N=42$) and $F160W$ centroids ($N=36$). Surprisingly however, no stacked signal is detected
regardless of the centroiding or smoothing method.
In Section \ref{sec:understanding_non_detections} we later eliminate the hypothesis that the non-detected sources are drawn from a different population to the 12 detections presented in Figures \ref{fig:gold_mosaic} and \ref{fig:silver_mosaic}, since they sample the same range of \MUV{}, \EWLya{}, \EWOIII{} and \dvLya{}, as shown in Figure
\ref{fig:fesc_populations}. Likewise the issue of spatial offsets e.g. LyC - $F160W$ inherent in any sample should not preclude a faint $F336W$ detection since the stacked $F336W$ signal is well-detected for the Gold and Silver subsamples (Figure \ref{fig:Det_stacks}).
The inevitable and remarkable conclusion, therefore, is that the mean $F336W$ signal in the non-detected sample must be uniformly much fainter than for the detected sample.
The $3\sigma$ upper limits for individual objects are typically $30.2$AB, and
quantitatively we can say that the $3\sigma$ upper limit we measure for the $F336W$ stack of $42$ non-detections is $31.8$AB. In addition to median stacks, we have also performed mean stacks of the images and they also show non-detections to a comparable $3\sigma$ upper limit.
Thus it appears the $F336W$ flux in our total sample is either detected individually or not at all. We defer discussion of this important result to Section \ref{sec:discussion}.

\subsubsection{Charge Transfer Efficiency and Stacked Non-Detections}

An important consideration for interpreting the lack of $F336W$ flux in the
stack of non-detections concerns the possible role of degraded Charge
Transfer Efficiency (CTE) in artificially suppressing flux from faint objects.
If charge was lost in the readout of the data owing to traps in the detector,
then one might expect suppression of very faint signals. Indeed, this physical
behavior is known to exist for the WFC3/UVIS detector and is actively mitigated
by the STScI pipeline, provided the electron background for the observations
is $\gtrsim12$ e- per pixel or charge packet.\footnote{See {\tt http://www.stsci.edu/hst/wfc3/ins\_performance/CTE/}.}.
Our program used post-flashing of the detector to reach
this background level, which should allow for the CTE corrections in the STScI
pipeline to recover efficiently the flux lost during readout. These CTE correction
tools are documented on the STScI website\footnote{See {\tt http://www.stsci.edu/hst/wfc3/tools/cte\_tools/}}.

However, the nominal flux limits reached by the stacks of individually non-detected
objects reaches to extremely faint limits ($\sim32$AB). Given the gain of the WFC3/UVIS
detector and the exposure time of our individual frames, we estimate that $1\sigma$
sources in the $F336W$ stack of non-detections would have a charge of only $\sim9$e- in
an individual exposure. According to the WFC3/UVIS CTE tools documentation, charge packets
of this size in a background of 12e- would lose $\sim20\%$ of their charge during
readout. To verify this expectation, we used
the {\tt wfc3uv\_cteforward} code by Jay Anderson to simulate the CTE effects on
charge packets of different amounts (1-1000e-) in differing backgrounds (0-15e-). We distribute
fake sources across a model of the WFC3/UVIS detectors and simulate the charge transfer
and readout process. We correct the expected flux in each simulated pixel for geometric
distortion applying relative exposure maps of the WFC3/UVIS detectora provided by STScI. We
use SEP to identify sources in the CTE-affected simulated images using the same
effective apertures used in our analysis, and then compute the fraction of original
charge lost during readout. We find that for 9-12e- sources, the CTE loss is 16-20\% for
12e- backgrounds. The CTE loss becomes extremely severe for $\lesssim5$e- backgrounds,
but such backgrounds are much lower than those in our observations.

For our stack of non-detections, this loss of charge could then
reduce a $1.25\sigma$ (11e- in a single exposure)
source below $1\sigma$ (9e- in a single exposure) flux levels. We verified this in further
simulations where we treated the loss of electrons from $9-15$e- sources owing to CTE
as a binomial process where the probability of survival for each electron was 80-84\%, using
$10^6$ realizations of their fluxes to determine how many sources could be pushed below $1\sigma$
in the final stack of non-detections. Importantly, no sources with charge packets ${\geq}12$e- in a single
exposure are expected to be suppressed below $1\sigma$ flux levels in the final stack of non-detections,
although individual (uncorrected for CTE) fluxes would be affected at the 20\% level. We therefore
conclude that for the backgrounds in our observations, CTE should not strongly influence
the stringent $F336W$ flux limits from our stack of non-detections.


\section{Analysis}\label{sec:analysis}

We now turn to using our $F336W$ detections and upper limits to derive the escape fraction \fesc{} of ionizing photons, both on a galaxy-by-galaxy basis for our sample and for the population as a whole. We likewise seek to correlate the escape fractions with our infrared spectroscopic measures of \OIII{} emission, primarily to test the hypothesis that a high escape fraction is connected with the intense \OIII{} emission that seems commonplace for star-forming sources in the reionization era.


\subsection{Relative Escape Fractions}\label{sec:f_esc_rel}

Estimating the escape fraction of Lyman continuum photons requires
knowledge about the intrinsic source spectrum before attenuation
by interstellar dust in the rest-ultraviolet or by the intergalactic medium blueward of Lyman-$\alpha$. We can use SED modeling to
constrain the escape fraction while simultaneously fitting for
other galaxy parameters on a source-by-source basis, and we
perform that analysis below. However, given the additional
uncertainties and model dependencies associated with SED fitting, we now consider
estimates of the escape fraction derived directly from the
source photometry.

The relative escape fraction of Lyman continuum photons, $f_{\mathrm{esc,rel}}(\mathrm{LyC})$, is often defined in terms of the
source flux $f_{900}$ at $\lambda_{\mathrm{rest}}=900$ \AA~and the
rest-ultraviolet flux $f_{1500}$ at $\lambda_{\mathrm{rest}}=1500$ \AA~
as
\begin{equation}
f_{\mathrm{esc,rel}}(\mathrm{LyC}) = \frac{(f_{900}/f_{1500})}{(L_{900}/L_{1500})t_{\mathrm{IGM}}},
\end{equation}
\noindent
where $(L_{900}/L_{1500})$ is the ratio of the intrinsic
spectrum at $\lambda_{\mathrm{rest}}=900$ \AA~and $\lambda_{\mathrm{rest}}=1500$ \AA~and $t_{\mathrm{IGM}}$ is the
transmission fraction in the Lyman continuum through the IGM \citep[e.g.,][]{steidel2001,inoue2005,shapley2006}.
Clearly both $(L_{900}/L_{1500})$ and $t_{\mathrm{IGM}}$ are
model-dependent quantities, and reflect assumptions about
the intrinsic stellar population spectrum and the range of
IGM absorption properties.

Here, we aim to provide an estimate of the relative escape fraction
that is easily interpreted before turning to a more sophisticated
estimate derived from the full set of photometric data. We
therefore define the estimated relative escape fraction
\begin{equation}
\label{eqn:tf_esc_rel}
\tilde{f}_{\mathrm{esc,rel}}(\mathrm{LyC}) = \frac{(f_{F336W}/f_{R})}{\min(L_{900}/L_{1500})\langle t_{\mathrm{IGM}}\rangle }.
\end{equation}
\noindent
Here we replace the measured flux ratio $(f_{900}/f_{1500})$
with our closest photometric flux ratio measure $(f_{F336W}/f_{R})$
using the $F336W$ and $R$ bands.
The quantity $\langle t_{\mathrm{IGM}}\rangle$ is the mean IGM transmission fraction
and, using the \citet{inoue2014} IGM absorption model, we find that
for our $z\sim3.1$ emitters $\langle t_{\mathrm{IGM}}\rangle\approx0.28$ averaged over the $F336W$ band.
We then need to define
the quantity $\min(L_{900}/L_{1500})$, the minimum intrinsic
luminosity density ratio expected for the source stellar populations.
We choose to use the minimum luminosity density ratio in Equation \ref{eqn:tf_esc_rel}
to provide upper limits on $\tilde{f}_{\mathrm{esc,rel}}(\mathrm{LyC})$
for our choice of $\langle t_{\mathrm{IGM}}\rangle$.
As we will see the estimated relative escape fraction can be
$\tilde{f}_{\mathrm{esc,rel}}(\mathrm{LyC})>1$, which physically
requires stellar populations with ages $t<10^{9}$ yr and/or a low opacity sight line through the IGM.

\begin{figure}
\includegraphics[width=3.5in]{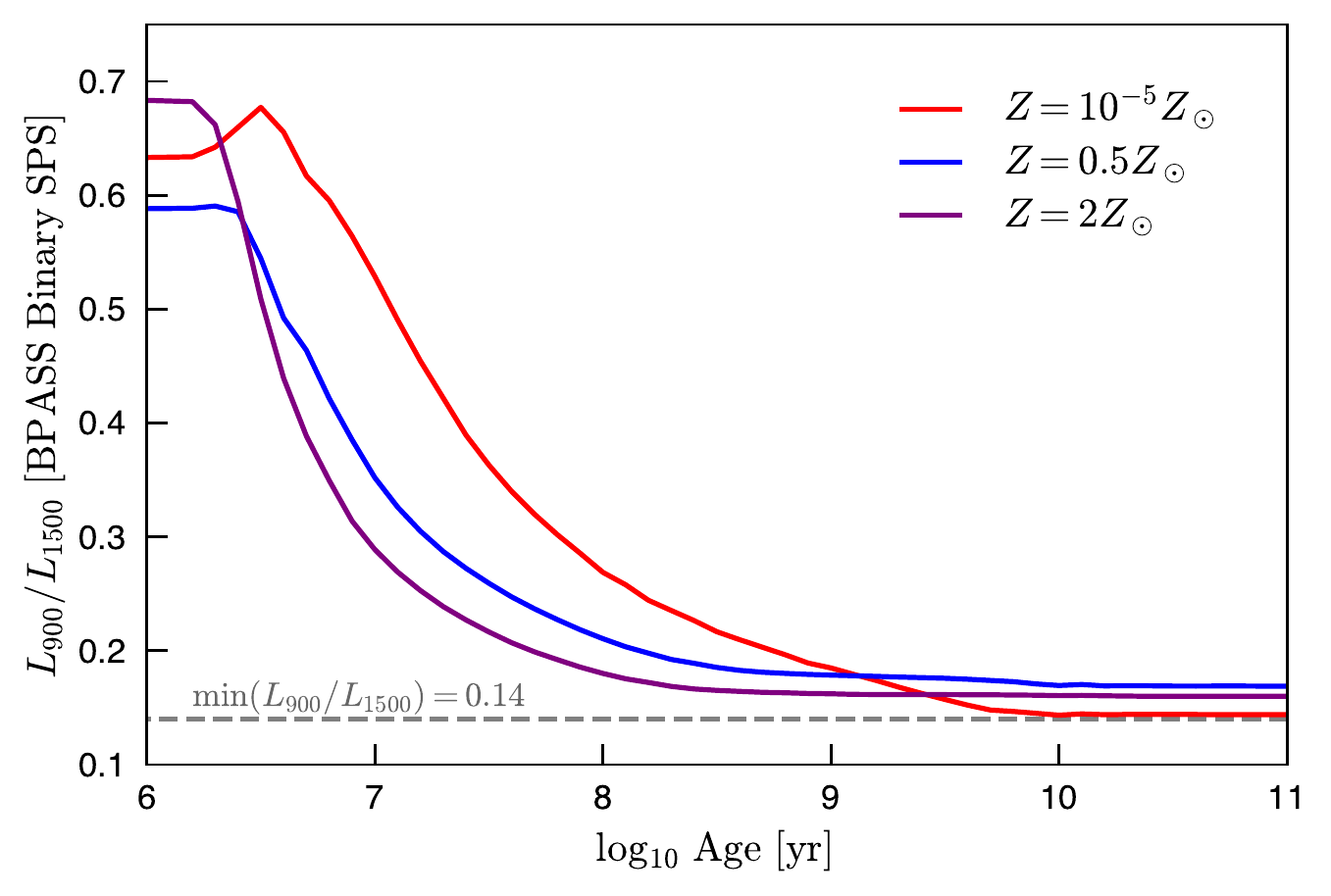}
\caption{The ratio $L_{900}/L_{1500}$ of the luminosity density at
900 \AA{} and 1500 \AA{} for constant star formation rate
BPASS binary stellar population models
as a function of age, for metallicities of $Z=10^{-5}Z_{\odot}$ (red),
$Z=0.5Z_{\odot}$ (blue), and $Z=2Z_\odot$ (purple). For
typical ages and metallicities of real galaxies, the luminosity
density ratio falls in the range $0.15\lesssim L_{900}/L_{1500}
\lesssim 0.3$. For the estimated relative escape fractions
inferred directly from the observed source flux ratios, we
will adopt a minimum ratio of $\min(L_{900}/L_{1500})=0.14$ (dashed
gray line).}
\label{fig:Lratio}
\end{figure}

Figure \ref{fig:Lratio} shows the intrinsic ratio of the Lyman
continuum and rest-UV luminosity densities for constant star
formation rate binary
stellar population models computing using
Version 2.1 of the Binary Population and Spectral Synthesis (BPASS) code \citep{eldridge2009a, eldridge2012a, stanway2016a, eldridge2017binary}. We compute the intrinsic ratio as a function
of stellar population age in the BPASS models, and plot the
quantity for metallicities $Z = [10^{-5}Z_{\odot},0.5Z_{\odot},2Z_{\odot}]$
assuming an upper stellar mass limit of $M = 100 M_{\odot}$.
For galaxies with stellar population ages older than $10^8$ years,
the typical ratio will be $0.15\lesssim L_{900}/L_{1500}
\lesssim 0.3$. Motivated by the behavior of these models, for our estimated relative escape fraction $\tilde{f}_{\mathrm{esc,rel}}$
we adopt the value
$\min(L_{900}/L_{1500})=0.14$.

\begin{figure}[t]
\includegraphics[width=3.5in]{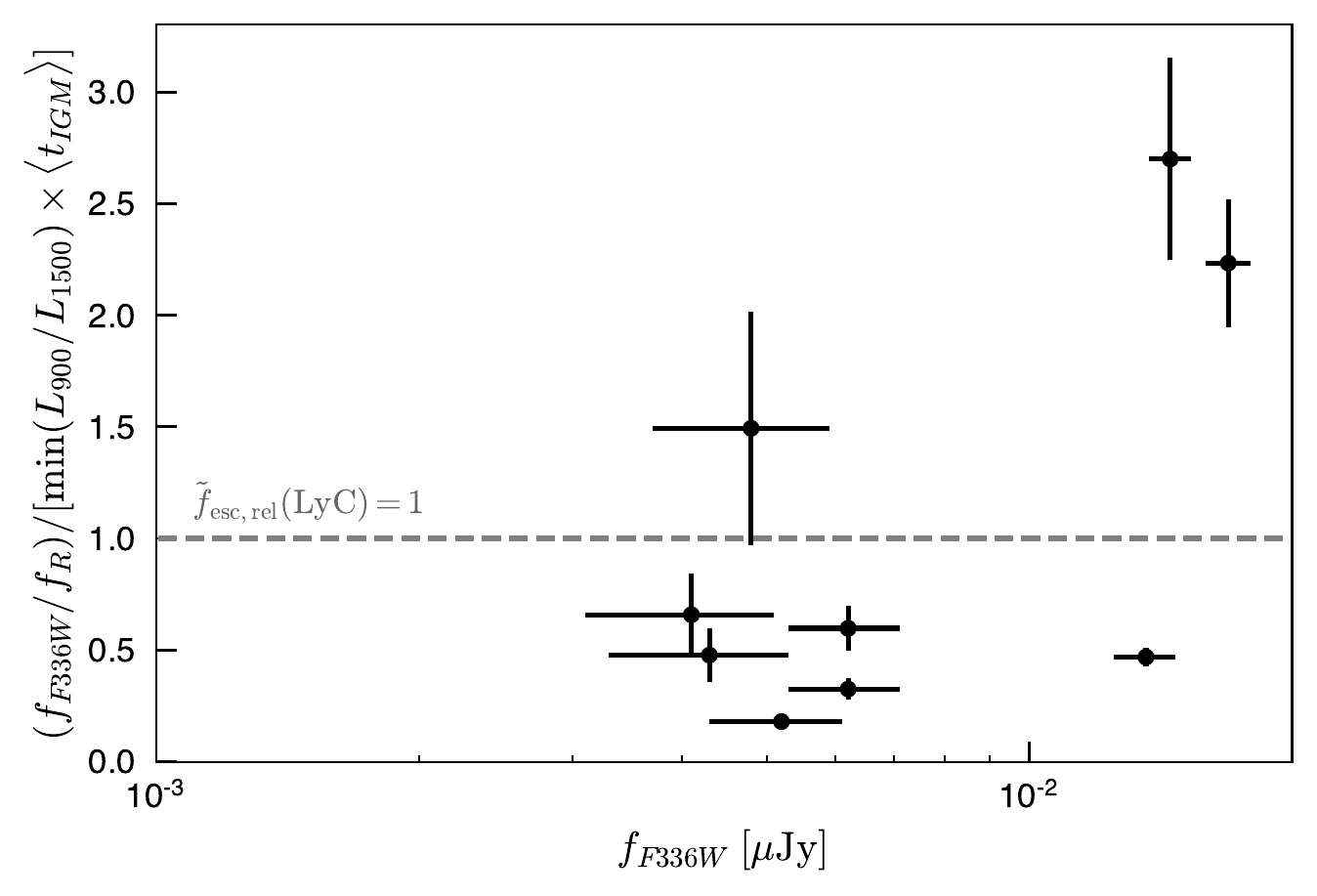}
\caption{Estimated relative escape fractions
$\tilde{f}_{\mathrm{esc,rel}}$ for our LAE sample with $R$ band detections, as
a function of the detected $F336W$ flux. Shown are the
flux ratios $f_{F336W}/f_R$ (black points) normalized by the minimum
intrinsic luminosity density ratio $\min(L_{900}/L_{1500})=0.14$
and the mean IGM transmission fraction
$\langle t_{\mathrm{IGM}}\rangle\approx0.28$. The vertical error bars indicate
the uncertainty on the estimated relative escape fractions
owing to the individual uncertainties
on the measured $F336W$ and $R$ band fluxes, while the
horizontal error bars correspond to the uncertainties on the
$F336W$ flux. For objects
above $\tilde{f}_{\mathrm{esc,rel}}=1$ (gray dashed line),
young stellar populations ($t\lesssim10^{9}$ yr) and/or
lower-than-average IGM absorption are required.
}
\label{fig:f_esc_rel}
\end{figure}

With concrete values for the quantities in the numerator of
Equation \ref{eqn:tf_esc_rel}, we can use the measured $F336W$
and $R$ band fluxes to estimate $\tilde{f}_{\mathrm{esc,rel}}$.
Figure \ref{fig:f_esc_rel} shows the estimated relative escape
fraction for our LAE sources as determined by the $F336W$ and
$R$ band flux ratios, assuming $\min(L_{900}/L_{1500})=0.14$
and $\langle t_{\mathrm{IGM}}\rangle\approx0.28$.
The uncertainties
on the estimated relative escape fraction
are computed by propagating the uncertainties on the measured
fluxes, and we include only sources with $R$ band detections.
The $\tilde{f}_{\mathrm{esc,rel}}$ values for individual sources
are listed in Table \ref{table:sed_params}, and fall in the range
$\tilde{f}_{\mathrm{esc,rel}}\approx0.18-2.7$.

\subsection{SED Model Fits}\label{sec:SEDs}
The absolute escape fraction is conventionally defined as the ratio of those LyC photons emerging compared to those intrinsic to the stellar population.
A first necessary step, therefore, is to determine the most likely stellar population and dust extinction for each source from which the intrinsic LyC radiation can be predicted.
Fortunately, the SSA22 sample has extensive multi-band photometry, and so the SEDs of many galaxies are well-constrained and provide the basis for this important step. A second requirement is to correct the detected $F336W$ flux upward to allow for line of sight absorption in the IGM. The mean IGM opacity increases as a function of redshift, and may vary between sources
owing to fluctuations in the number of absorbers along the line of sight.

We begin by fitting the SEDs of all the sources in our sample, regardless of whether they have $F336W$ detections. The SED data comprises $U$-band data from the Canada-France-Hawaii telescope, $B$, $V$, $R$, $i^\prime$ and $z^\prime$ from Subaru, $J$ and $K$ from UKIRT as well as $F160W$ from HST and Channel 3 and 4 coverage from IRAC on-board the Spitzer Space Telescope. These precursor datasets are summarized in Table \ref{table:opt_nir_photometry} and the individual photometry is shown in Table \ref{tbl:photometry}.

We use BPASS v2.1
to generate synthetic spectral energy distributions that we fit to the data\footnote{Although one of our Gold subsample objects (86861) is a weak LAE-AGN, from here onwards we proceed in using the BPASS models for consistency with the rest of our LAEs.}. Assuming a constant star formation history, $Z=0.1Z_{\odot}$ metallicities, and stellar masses in the range $M\in[0.1,100]M_{\odot}$\footnote{Models with a $300M_{\odot}$ cut-off increase the ionizing flux by $\sim 5\%$ \citep{eldridge2017binary} given the same rest-frame UV luminosity. The choice of model will therefore introduce a $\sim 10\%$ uncertainty in \fesc{}.}, we couple the BPASS models with the {\it MULTINEST} \citep{feroz2008a,feroz2009a} nested sampler to perform Bayesian parameter estimation on each galaxy's star formation rate ($A_{\mathrm{SFR}}\in[0,100]M_{\odot}/\mathrm{yr}$), stellar age ($t_{\mathrm{age}}\in[0,t_{\mathrm{max}}(z)]$),
dust extinction assuming the \citet{gordon2003a} Small Magellanic Cloud reddening law ($E(B-V)\in[0,1])$,
and the escape fraction ($f_{\mathrm{esc}}\in[0,1]$). When performing parameter estimation, we use flat priors for all parameters.
When fitting the BPASS SEDs to the photometry we have examined both single and binary star stellar populations, and report results for binary population models since these fits produce conservatively lower inferred \fesc{}.
Nebular continuum and line emission is included following the precepts of
\citet{robertson2010a}, with the strength of the nebular emission scaling with the Lyman continuum photon production rate and moderated by the escape fraction. Attenuation from the intergalactic medium owing to neutral hydrogen absorption is included following \citet{inoue2014}, and is applied according to the spectroscopic redshift of each source (or the narrow band \Lya{} redshift if spectroscopy was unavailable). When fit, the escape fraction simply adjusts the model $F336W$ flux by a multiplicative factor and we incorporate any possible dust attenuation of the Lyman continuum into the value of $f_{\mathrm{esc}}$. However, we do limit $f_{\mathrm{esc}}$ to be less or equal to the
transmission permitted by the dust attenuation expected for a given $E(B-V)$.
Model photometry is calculated from the model spectra following \citet{papovich2001a}.

\begin{figure*}
\centering
\includegraphics[width=\linewidth]{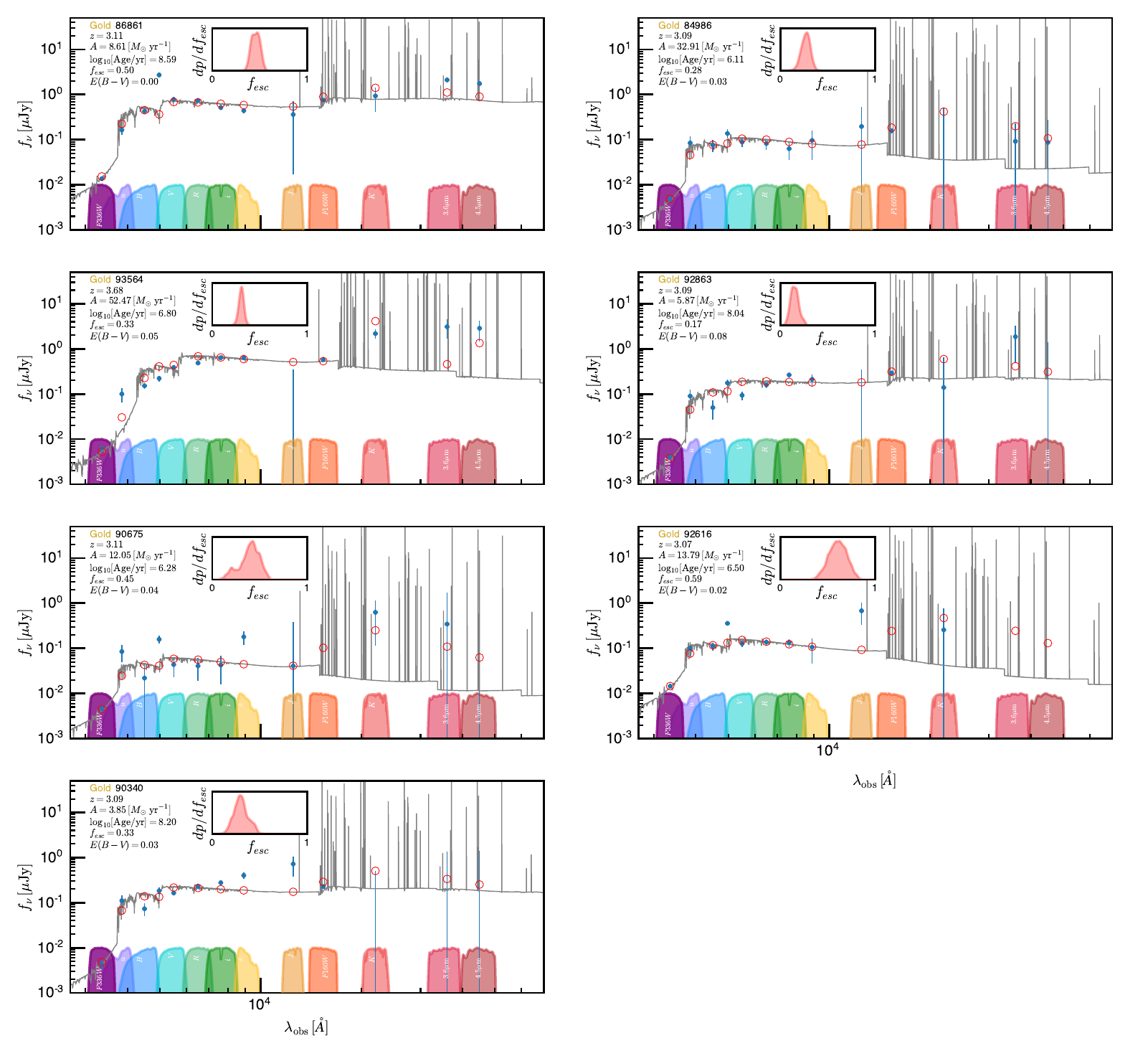}
\caption{SED fits to Gold subsample $z\simeq 3.1$ LAEs. The photometric
data (blue points with error bars) for each source across 12 bands (colored regions) is used to constrain the
SED model fit (gray line), resulting in the model photometry (open red circles).
The maximum likelihood parameters
for the star formation rate, the age of the constant
star formation rate stellar population, the extinction,
and the Lyman continuum escape fraction are reported. The insets show the marginal constraint on the
escape fraction $f_{esc}$ for each object.}
\label{fig:sed_Gold}
\end{figure*}

\begin{figure*}
\centering
\includegraphics[width=\linewidth]{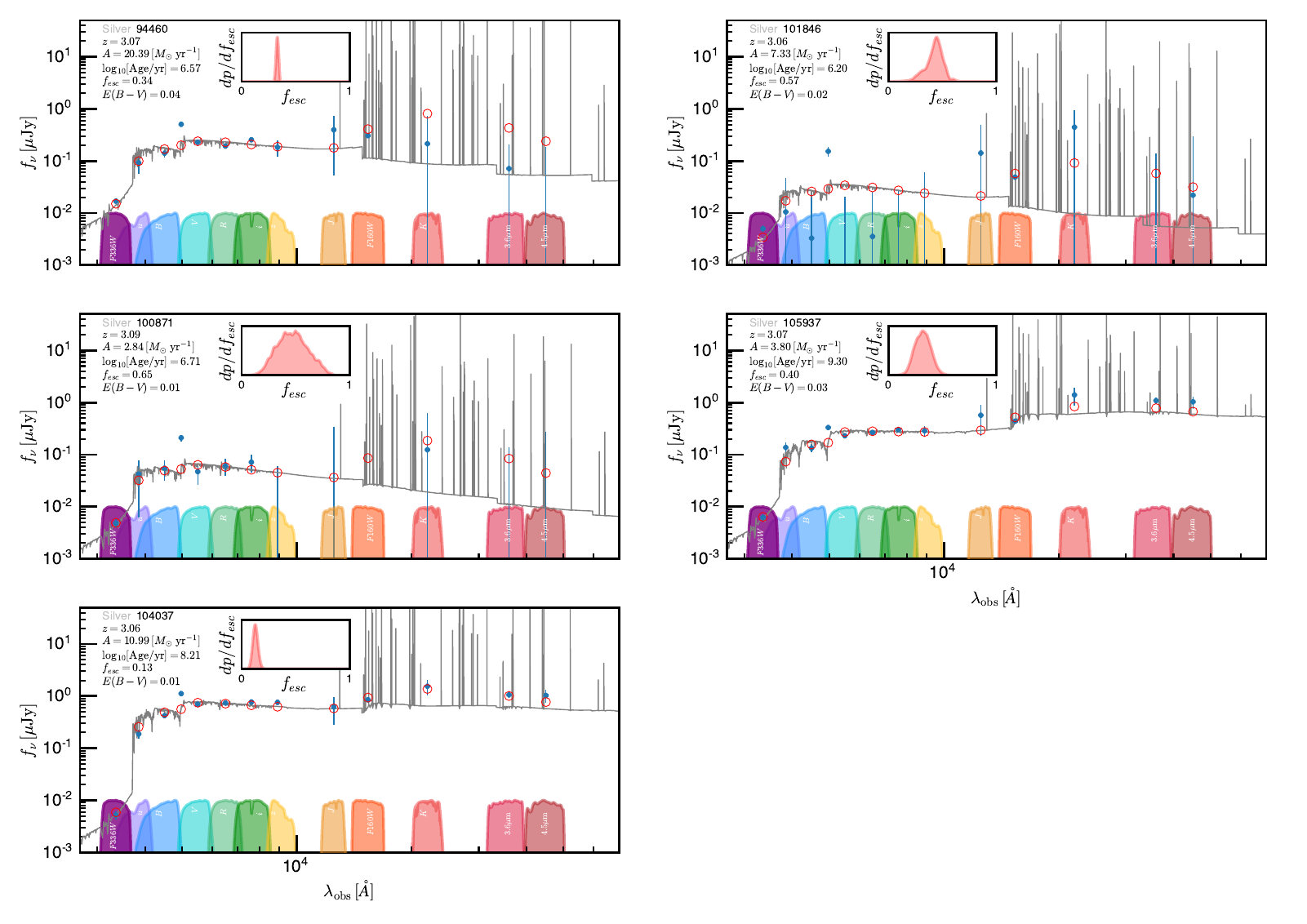}
\caption{SED fits to Silver subsample $z\simeq 3.1$ LAEs. The photometric
data (blue points with error bars) for each source across 12 bands (colored regions) is used to constrain the
SED model fit (gray line), resulting in the model photometry (open red circles).
The maximum likelihood parameters
for the star formation rate, the age of the constant
star formation rate stellar population, the extinction,
and the Lyman continuum escape fraction are reported. The insets show the marginal constraint on the
escape fraction $f_{esc}$ for each object.}
\label{fig:sed_Silver}
\end{figure*}

Figures \ref{fig:sed_Gold} and \ref{fig:sed_Silver} show model SED fits to the individual
Gold and Silver subsample objects.
For each object, the maximum likelihood model parameters for star formation rate, stellar age, $E(B-V)$,
and escape fraction are indicated. Inset panels in the figures indicate the marginal distribution for
$f_{esc}$ determined from each model fit. The SED parameter constraints are recorded in Table \ref{table:sed_params},
which lists the mean and $1\sigma$ width of the posterior distributions\footnote{The error reported here is inferred from the posterior distribution, but does not include systematic effects associated with model uncertainties.}.
The quality of the fits vary depending on the photometric constraints available for each object, and the
constraints on $f_{esc}$ vary correspondingly.
The typical escape fractions
inferred from individual SED fits
are $f_{esc}\approx0.4$ for the
Gold and Silver subsamples, with substantial spread.

\subsubsection{SED Model Tests}
\label{sec:SED_tests}

The SED modeling provides stellar population constraints on the objects, and enables a model-dependent inference of
the escape fraction $f_{esc}$. While the details of the model do not change whether Lyman continuum flux is detected
in our sample and could not permit a conclusion that $f_{esc}\sim 0$, assessing the influence of our model assumptions
on our derived parameters is important. We consider some important potential issues below.

{\it Rest-Frame UV and Optical Photometry:} The rest-frame optical photometry provides constraints on the presence of evolved
stellar populations and, for some combinations of redshifts and photometric bands, the possible influence
of nebular line emission. Without constraints on the escaping Lyman continuum flux, the permitted
contribution of internally absorbed Lyman continuum photons to the nebular emission can vary widely. Whether the $F336W$
or IRAC fluxes influence the stellar population parameters therefore depend on the detailed shape of the object SED.
For instance, Gold Object 86861 has a model escape fraction of $f_{esc} = 0.46\pm {0.05}$ when including all the photometric
data. The rest-frame UV, $F160W$, and IRAC data for this object provide reasonably tight constraints on the object parameters,
such that if the $F336W$ data is removed from the fit, the star formation rate and age of the object only change less than $10\%$
to $A_{\mathrm{SFR}} = 8.3~M_{\odot}/\mathrm{yr}$ and $\log_{10} t_{\mathrm{age}} = 8.9$.
However, the IRAC data for this object does help resolve the age-SFR degeneracy and removing the IRAC data decreases the age
to $\log_{10} t_{\mathrm{age}} = 8.3$, increases the star formation rate to $A_{\mathrm{SFR}} = 9.6~M_{\odot}/\mathrm{yr}$,
and decreases the inferred escape fraction to $f_{esc} = 0.38$. For other objects with good photometric constraints in the
rest frame UV and at $F160W$, such as Silver Object 104037, the escape fraction constraints can change by less than $20\%$ when
the IRAC photometry is ignored.

{\it IGM Absorption:} Without considering additional possible constraints on the ionizing emissivity of the LAEs,
the escape fraction inferred by the SED modeling will directly anti-correlate with the IGM attenuation along the
line of sight to any object. Models of the IGM absorption by \citet{madau1995a} or \citet{inoue2014} connect the
IGM absorption with the occurrence of neutral hydrogen systems along the line of sight, and variations in the
absorption to the statistical variance of these absorbers \citep[e.g.,][]{inoue2008a}. Since the escape fraction
is bounded in the range $f_{esc}\in[0,1]$ and the IGM absorption depends exponentially on the line-of-sight opacity,
the detection of the Lyman continuum in multiple objects may suggest that our SSA22 line-of-sight has lower
than average opacity.
If the IGM transmissivity is higher than the average
$\langle t_{\mathrm{IGM}} \rangle \approx 0.3$ we assume,
then our inferred escape fractions could go down
by a factor of two at most. Given that we find a substantial spread in the inferred $f_{esc}$ for our objects, we
can only conclude that the IGM transmissivity is not uniformly low.

{\it Stellar Population Binarity:} We assume binary stellar populations in the BPASS models. For a constant star
formation rate population with an age $t_{\mathrm{age}}>100~\mathrm{Myr}$, the difference in the Lyman continuum
flux per unit UV luminosity density is only
$\Delta \log_{10} \xi_{ion,0} \sim 0.05$ for $Z=Z_{\odot}/10$ and the difference
overall production rate of Lyman continuum photons is $\Delta \log_{10} N_\mathrm{ion} \sim 0.1$. The differences
in $f_{esc}$ inferred from changing between single and binary stellar populations therefore vary less than
the typical uncertainties associated with star formation history, dust, and age.

{\it Dust Model:} Variations in the dust model can influence the escape fraction
inferred from SED modeling, as the absorption in the rest-frame UV for a given
$E(B-V)$ can differ and result in different ratios between the intrinsic model and
observed Lyman continuum fluxes. Many of our objects are very blue and permit only
very small values of $E(B-V)$.
For instance, our Gold subsample object 92863 has an inferred SMC $E(B-V)=0.07$.
Using a \citet{calzetti1994a} dust law results in an $E(B-V)=0.07$, but the best-fit
star formation rate has declined by $40\%$ to $A_{\mathrm{SFR}}=3.1~M_{\odot}/\mathrm{yr}$,
the age has increased to $\log_{10} t_{\mathrm{age}} = 9.3$, and the inferred
escape fraction increases substantially to $f_{\mathrm{esc}} = 0.39$. However, these changes
represent only $\sim2\sigma$ changes compared to the SMC dust-based model
parameter constraints, and while objects with non-negligible dust may have a systematic
uncertainty associated with the dust model their SED fits tend to be less constrained
anyway. Fortunately, our sample of intrinsically blue LAEs will suffer less from the
systematic uncertainties associated with dust than Lyman continuum surveys of more evolved
LBG samples.

\begin{table*}
\centering
\caption{SED Parameter Constraints}
\label{table:sed_params}
\renewcommand{\arraystretch}{1.4}
\begin{tabular}{@{}lcccccc@{}}
\hline
\hline
ID & $A_{\mathrm{SFR}}$ [$M_{\odot}$/yr$^{-1}$] & $t_{\mathrm{age}}$ [$\log_{10}$ yr] &  $M_\star$ [$\log_{10} M_\odot$] & $E(B-V)$  & $f_{\mathrm{esc}}$ & $\tilde{f}_{\mathrm{esc,rel}}$ \\
\hline
\multicolumn{7}{c}{Gold Sample}\\\hline
86861* & $9.06\pm{0.42}$ & $8.49\pm{0.10}$ & $9.44\pm{0.08}$ & ${<}0.003$ & $0.46\pm{0.05}$ & $0.47\pm0.04$ \\
93564 & $89.86\pm{70.54}$ & $6.68\pm{0.27}$ & $8.54\pm{0.07}$ & $0.050\pm{0.004}$ & $0.31\pm{0.03}$ & $0.32\pm0.05$ \\
90675 & $12.53\pm{6.47}$ & $6.34\pm{0.20}$ & $7.38\pm{0.12}$ & $0.03\pm{0.02}$ & $0.39\pm{0.11}$ & -- \\
90340 & $4.29\pm{0.77}$ & $7.90\pm{0.28}$ & $8.52\pm{0.22}$ & $0.03\pm{0.01}$ & $0.30\pm{0.08}$ & $0.48\pm0.12$ \\
84986 & $19.39\pm{7.83}$ & $6.32\pm{0.22}$ & $7.57\pm{0.10}$ & $0.03\pm{0.01}$ & $0.26\pm{0.05}$ & $1.49\pm0.52$ \\
92863 & $7.75\pm{6.77}$ & $7.78\pm{0.60}$ & $8.61\pm{0.43}$ & $0.07\pm{0.01}$ & $0.15\pm{0.04}$ & $0.66\pm0.18$ \\
92616 & $26.13\pm{11.15}$ & $6.31\pm{0.16}$ & $7.68\pm{0.07}$ & $0.02\pm{0.01}$ & $0.60\pm{0.09}$ & $2.70\pm0.45$ \\
\hline
\multicolumn{7}{c}{Silver Sample}\\\hline
94460 & $23.81\pm{2.00}$ & $6.51\pm{0.03}$ & $7.89\pm{0.01}$ & $0.036\pm{0.003}$ & $0.33\pm{0.02}$ & $2.23\pm0.29$ \\
105937 & $4.91\pm{1.44}$ & $8.95\pm{0.30}$ & $9.64\pm{0.22}$ & $0.04\pm{0.01}$ & $0.32\pm{0.07}$ & $0.60\pm0.10$ \\
104037 & $10.94\pm{1.04}$ & $8.22\pm{0.15}$ & $9.26\pm{0.11}$ & $0.011\pm{0.004}$ & $0.13\pm{0.02}$ & $0.18\pm0.03$ \\
101846 & $4.92\pm{2.73}$ & $6.47\pm{0.24}$ & $7.08\pm{0.13}$ & $0.03\pm{0.04}$ & $0.42\pm{0.09}$ & -- \\
100871 & $11.00\pm{6.05}$ & $6.37\pm{0.22}$ & $7.33\pm{0.12}$ & $0.02\pm{0.02}$ & $0.47\pm{0.14}$ & -- \\
\hline
\multicolumn{7}{c}{Composite SEDs}\\\hline
Gold & $6.50\pm{1.15}$ & $7.52\pm{0.27}$ & $8.32\pm{0.21}$ & $0.059\pm{0.005}$ & $0.22\pm{0.03}$ & $0.83\pm0.05$ \\
Silver & $3.10\pm{0.57}$ & $8.97\pm{0.26}$ & $9.46\pm{0.20}$ & $0.01\pm{0.01}$ & $0.51\pm{0.08}$ & $0.73\pm0.05$ \\
Non-Detections& $1.51\pm{0.14}$ & $8.28\pm{0.17}$ & $8.46\pm{0.14}$ & $0.005\pm{0.004}$ & ${<}0.005$ & ${<}0.006$ \\
\hline
\end{tabular}

\\
\vspace{-1mm}
\begin{flushleft}
\scriptsize
The ``--'' denotes objects with no $R$-band detection, and correspondingly no measure of $\tilde{f}_{\mathrm{esc,rel}}$.
\end{flushleft}
\end{table*}


\subsection{Measured vs. Model Line Fluxes}
\label{sec:model_line_flux}

The SED models are fit to the observed photometric fluxes
for each source, but our spectroscopic campaign has also provided independent measures of
the fluxes of \Hb{}, and \OIII{} that can be used to assess
the validity of the model line emission that we incorporate in
the rest-frame optical source SED.
In the case of \Hb{}, examining the ratio of the model line flux to that
observed for a range of the Lyman continuum photon production rate,
we find $\langle f_{mod}/f_{obs} \rangle = 1.44\pm1.81$.
Since the model line fluxes depend on $(1-f_{esc, \mathrm{SED}})$ determined from the SED fit,
this illustrates the degree of self-consistency
between the observed and modeled Lyman continuum flux, the
inferred escape fractions $(1-f_{esc,\mathrm{SED}})$, and
the method for computing the model line strengths that
contribute to the photometric data.

\subsection{Composite SEDs}

The photometry of our sources derives from a combination of
ground and space-based imaging, with a range of sensitivity and
spatial resolution. Excepting $F336W$, the rest-frame UV measurements all come from
ground-based data. While this data is of high quality, for
some objects it permits a range of stellar population parameters
that provide statistically similar model fits. Various
combinations of star formation rate and age can produce the
same rest-UV flux given these uncertainties, but would lead
to a range of inferred $f_{esc}$ as illustrated by the
marginal distributions shown in Figures \ref{fig:sed_Gold} and \ref{fig:sed_Silver}.
Given the homogeneity
of our sample objects and their comparable redshifts, we have constructed
composite photometry for the Gold, Silver, and $F336W$ Non-Detection subsamples
and performed SED model fits to the composite data.

The composite SEDs were generated by first cutting out $6\times 6$ arcsecond
postage stamps,
centered on the Subaru positions for each of the LACES LAEs, for all of the
available photometric bands. The composites were generated by stacking
these cutout images for each band for the Gold, Silver and non-detected subsamples
using the IRAF task {\tt imcombine}. Only LAEs within the SSA22 protocluster
at approximately $z\simeq 3.1$ were included and LAEs with $z>3.1$, such as ID 93564 were
excluded from the composites.

The errors for the composite SEDs were calculated by first masking
all objects with $\rm SNR>3$ in each image. Regions were then selected
from the remaining noise, ensuring no overlap with masked areas of the image.
These regions were then used to make stacked images of the noise,
stacking the same number of images used in the Gold,
Silver, and non-detection composites ($\rm N=6, 5, 42$ respectively).
This process was repeated across every photometric band for 500 stacks
of the noisy regions and the $1\sigma$ upper limits were calculated.

Table \ref{table:composite} provides the flux density and associated
$1\sigma$ uncertainty measured in each
band for the composite stacks for our Gold, Silver, and non-detection subsamples.
Using the higher-precision composite SEDs, we again perform our SED model fits
to explore possible inferred differences between the composite properties
of our subsamples.

\begin{table}
\centering
\caption{Composite SEDs}
\label{table:composite}
\renewcommand{\arraystretch}{1.4}
\begin{tabular}{@{}lccc@{}}
\hline
\hline
 & Gold Sample & Silver Sample & Non-detection Sample  \\
Band & Flux [nJy] & Flux [nJy] & Flux [nJy]\\
\hline
$F336W$        	& $   5.29 \pm  0.25 $ & $  5.86 \pm  0.25 $ & $ <0.24 $ \\
$u$          	& $   99.7 \pm  15.7 $ & $ 107.7 \pm  17.2 $ & $  26.2 \pm   7.1 $ \\
$B$          	& $   85.9 \pm   9.2 $ & $ 130.9 \pm   8.5 $ & $  67.4 \pm   6.0 $ \\
NB497      		& $  226.8 \pm  14.6 $ & $ 370.4 \pm  10.6 $ & $ 259.3 \pm   6.4 $ \\
$V$            	& $  135.2 \pm  10.2 $ & $ 217.8 \pm   8.1 $ & $ 110.3 \pm   6.4 $ \\
$R$            	& $  162.8 \pm   7.8 $ & $ 205.5 \pm   9.6 $ & $ 110.2 \pm   5.3 $ \\
$i^{\prime}$    & $  194.0 \pm  11.3 $ & $ 257.0 \pm  12.0 $ & $ 112.6 \pm   6.6 $ \\
$z^{\prime}$    & $  216.9 \pm  20.5 $ & $ 196.6 \pm  23.2 $ & $ 120.1 \pm  11.3 $ \\
$J$            	& $  381.2 \pm 189.4 $ & $ 292.2 \pm 197.8 $ & $ 175.3 \pm  78.9 $ \\
$F160W$        	& $  267.6 \pm  16.3 $ & $ 306.2 \pm  16.0 $ & $ 111.9 \pm   6.9 $ \\
$K$            	& $  248.2 \pm 301.8 $ & $ 809.6 \pm 326.2 $ & $ 123.8 \pm 120.4 $ \\
$[3.6\mu m]$	& $  654.8 \pm 150.5 $ & $ 591.3 \pm  60.6 $ & $ 340.3 \pm  72.1 $ \\
$[4.5\mu m]$	& $  764.0 \pm 213.6 $ & $ 462.5 \pm 130.1 $ & $ 286.1 \pm  50.8 $ \\
\hline
\end{tabular}

\end{table}

\subsubsection{Gold Subsample Composite SED}

\begin{figure}
\centering
\includegraphics[width=\linewidth]{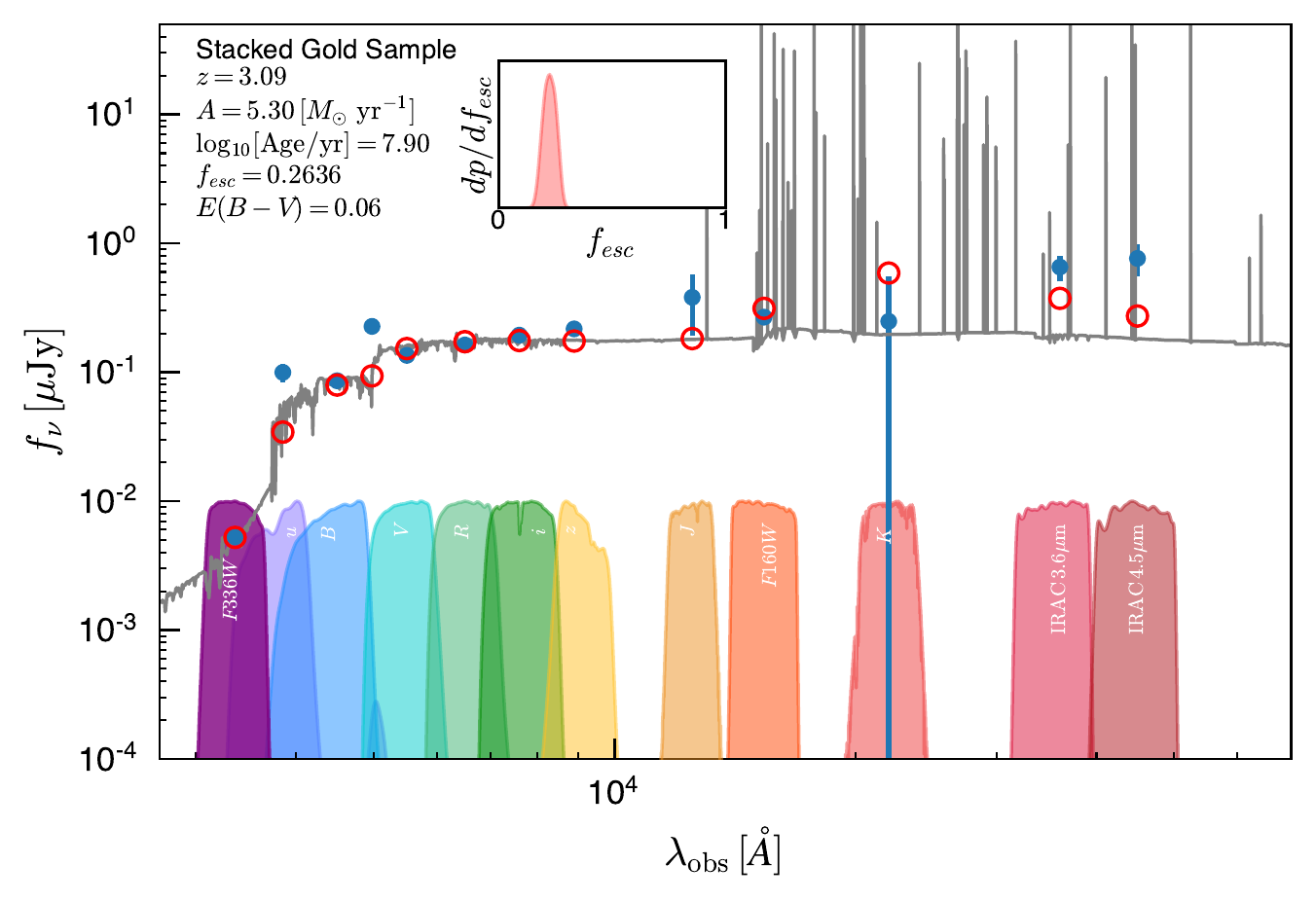}
\caption{SED model fit to photometry of stacked Gold sample detections.}
\label{fig:gold_stacked_sed}
\end{figure}

Figure \ref{fig:gold_stacked_sed} shows the maximum likelihood
model fit to the stacked photometry of the Gold
sample. This composite at $z\approx3.1$ appears to be consistent with
a very young star-forming population, a reddening of
$E(B-V)\approx0.06$, and
an escape fraction of $f_{\rm esc}\approx 0.2$.
This SED-inferred
escape fraction is lower than the relative escape fraction
suggested by ratio of the 900 \AA~ and 1500 \AA~ rest-frame flux densities and
the typical \fesc{} suggested by the individual
Gold subsample object fits.
This discrepancy results from improved constraints on the
rest-frame UV portion of the spectrum gained by stacking. Compared to the other composites discussed below, the mean age is younger although less certain. This reflects the wider range of inferred ages in the individual objects in the Gold sample as listed in Table \ref{table:sed_params}. If the mean age were increased, making it more consistent with that for the other composites, the inferred escape fraction for the Gold sample would be larger. We note that the SED model does not fit the IRAC data particularly well, but for the composite
Gold sample photometry ignoring the IRAC data entirely changes the maximum likelihood parameters by $<10\%$ fractionally.
For the escape fraction, fitting the composite Gold sample with or without including the IRAC data
changes the maximum likelihood fit by only $\Delta f_{esc}=0.01$.

\subsubsection{Silver Subsample Composite SED}

\begin{figure}
\centering
\includegraphics[width=\linewidth]{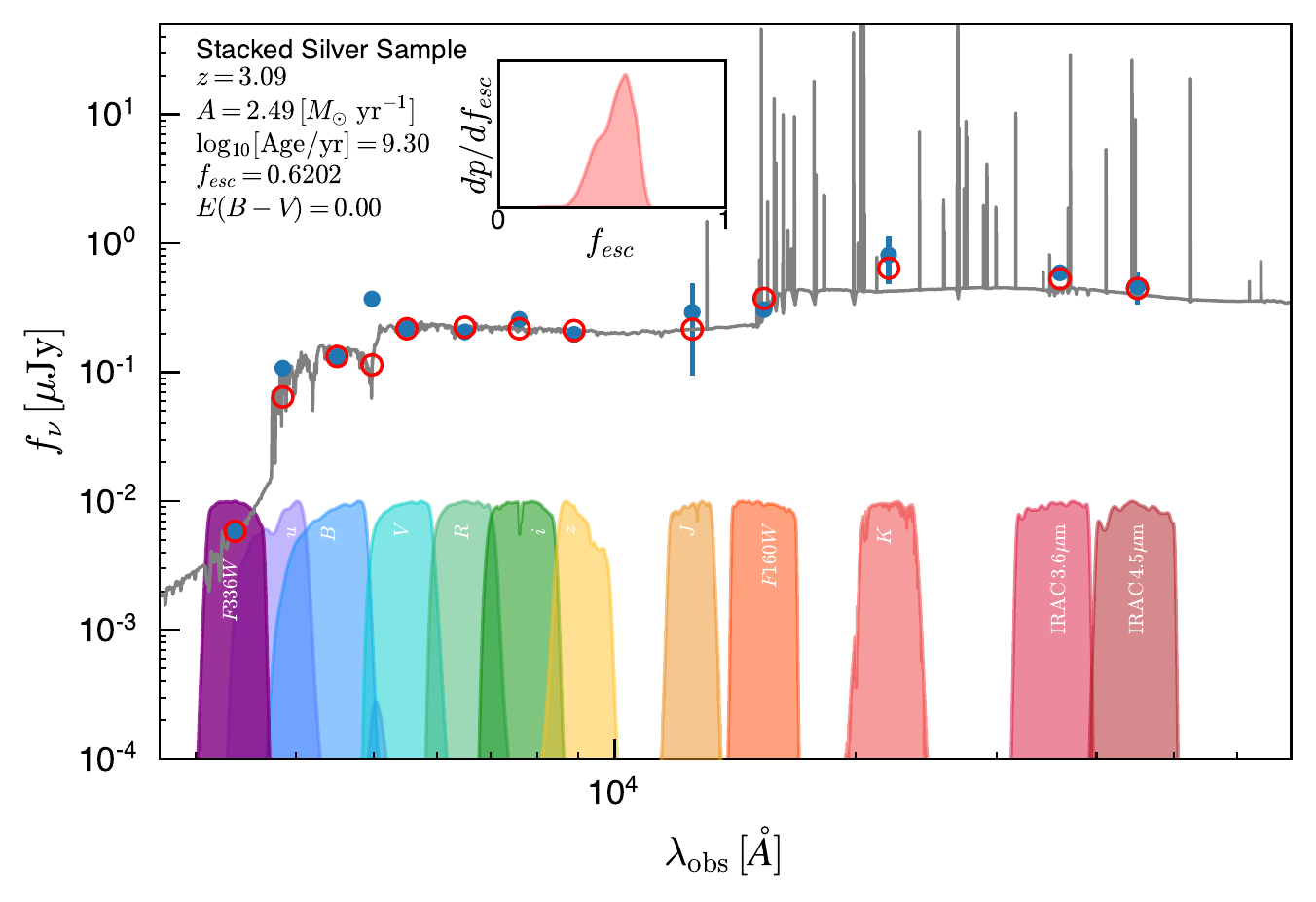}
\caption{SED model fit to photometry of stacked Silver sample detections.}
\label{fig:silver_stacked_sed}
\end{figure}

Figure \ref{fig:silver_stacked_sed} shows the maximum likelihood
model fit to the stacked photometry of the Silver
subsample. Relative to the Gold subsample composite, the Silver
subsample composite at $z\approx3.09$ is consistent with
being older and forming stars at a lower rate
and with a lower reddening $E(B-V)\approx0.01$. The inferred
Silver composite escape fraction is higher than for the Gold composite,
with $f_{esc}\approx0.5$.
How representative the escape fraction inferred for
this composite Silver SED is for the Silver sample
objects individually is questionable, as the photometry for the individual
objects varies widely. The stacked Lyman continuum flux is heavily influenced by
94460, but this object is faint in IRAC and individually has a very young inferred
age with moderate escape fraction. The objects 104037 and 105397 contribute greatly
to the rest-frame optical emission of the composite Silver SED, but these
objects are fairly bright and one is inferred to be old. The combination
of these SEDs leads to an old composite age with a necessarily large escape
fraction. Clearly some caution is warranted in generalizing the results of
the composite Silver SED fit to all LAEs in our sample.

\subsubsection{$F336W$ Non-Detection Subsample Composite SED}

\begin{figure}
\centering
\includegraphics[width=\linewidth]{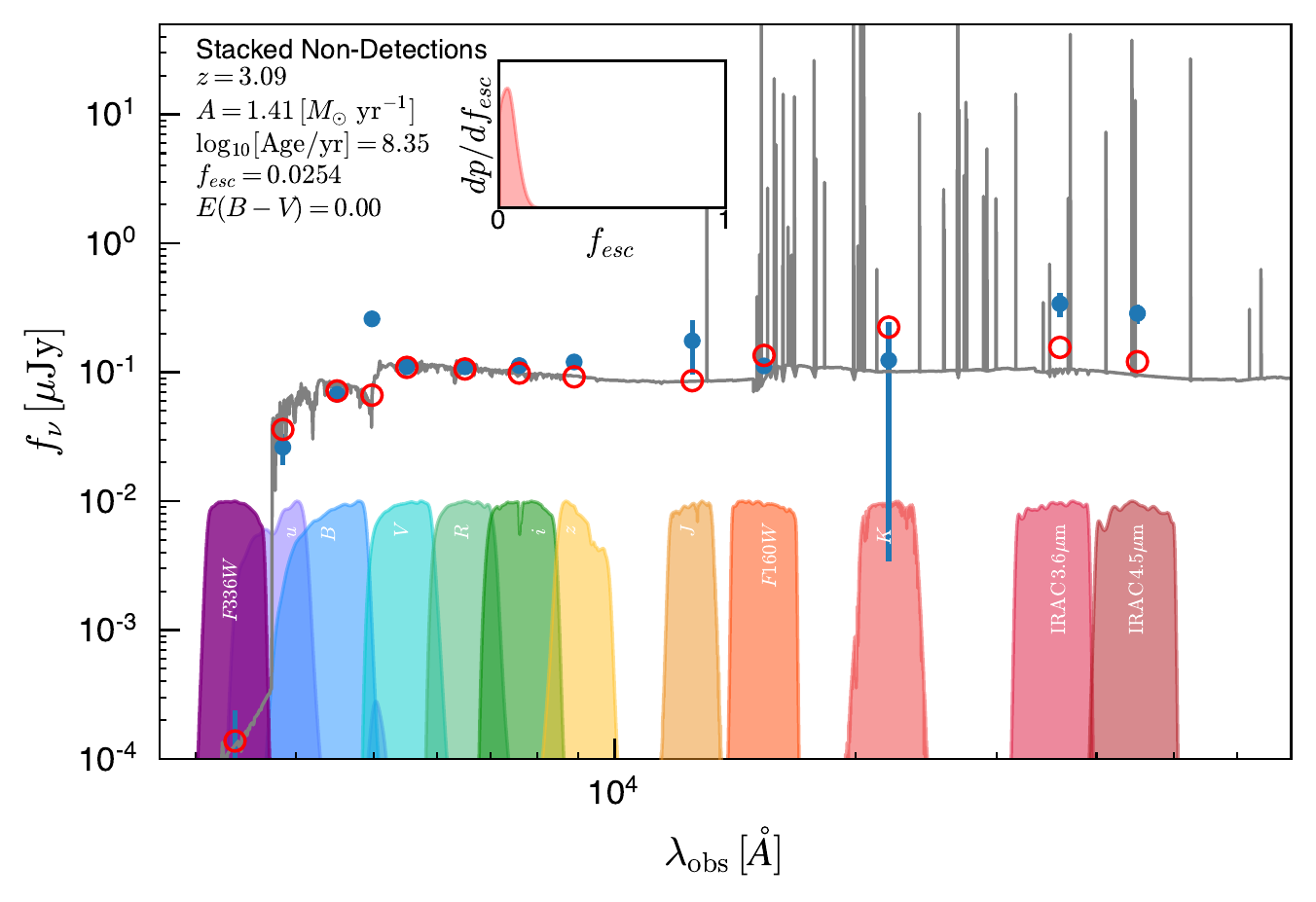}
\caption{SED model fit to photometry of stacked non-detections.}
\label{fig:nd_stacked_sed}
\end{figure}

Figure \ref{fig:nd_stacked_sed} shows the maximum likelihood
model fit to the stacked photometry of the $F336W$ Non-Detection
subsample. This composite at $z\sim3.1$ is consistent with a
500 Myr-old stellar population forming stars at $A_{\mathrm{SFR}}\lesssim~2M_{\odot}/\mathrm{yr}$,
lightly reddened, and with a Lyman escape fraction close to zero (i.e., $f_{esc}<0.005$).
In our models, the low escape fraction results in strong nebular
continuum and line emission in the rest-frame optical, and the model
photometry agrees well with the stacked photometric data at these
wavelengths.

\begin{table*}
\centering
\caption{Photometry of the LyC Leaking Candidates}
\label{tbl:photometry}
\renewcommand{\arraystretch}{1.25}
\begin{tabular}{@{}lcccccccccccc@{}}
\hline
ID & $u$ & $B$ & NB$497$ & $V$ & $R$ & $i^{\prime}$ & $z^{\prime}$ & $J$ & F$160$W & $K$ & $[3.6]$ & $[4.5]$\\
\hline
\multicolumn{13}{c}{Gold Sample}\\
\hline
$86861^{*}$ & $25.9$ & $24.8$ & $22.8$ & $24.2$ & $24.2$ & $24.6$ & $24.8$ & ${>}23.9$ & $24.2$ & ${>}23.4$ & $23.1$ & $23.3$ \\
$93564$ & ${>}26.3$ & $25.9$ & $25.5$ & $24.9$ & $24.7$ & $24.4$ & $24.4$ & ${>}23.9$ & $24.5$ & $23.1$ & ${>}22.4$ & ${>}22.3$ \\
$90675$ & ${>}26.3$ & ${>}26.8$ & $25.9$ & ${>}26.9$ & ${>}26.9$ & ${>}26.6$ & ${>}25.7$ & ${>}23.9$ & $25.5$ & ${>}23.4$ & ${>}22.4$ & ${>}22.3$ \\
$90340$ & $26.3$ & $26.7$ & $25.7$ & $25.9$ & $25.5$ & $25.3$ & $24.9$ & ${>}23.9$ & $25.4$ & ${>}23.4$ & ${>}22.4$ & ${>}22.3$ \\
$84986$ & ${>}26.3$ & $26.7$ & $26.1$ & $26.5$ & $26.6$ & ${>}26.6$ & ${>}25.7$ & ${>}23.9$ & $25.8$ & ${>}23.4$ & ${>}24.8$ & ${>}24.5$ \\
$92863$ & ${>}26.3$ & ${>}26.8$ & $25.8$ & $26.5$ & $25.9$ & $25.3$ & $25.6$ & ${>}23.9$ & $25.3$ & ${>}23.4$ & ${>}22.4$ & ${>}22.3$ \\
$92616$ & ${>}26.3$ & $26.3$ & $25.0$ & $26.1$ & $26.1$ & $26.1$ & ${>}25.7$ & ${>}23.9$ & $--$ & ${>}23.4$ & ${>}24.8$ & ${>}24.1$ \\
\hline
\multicolumn{13}{c}{Silver Sample}\\
\hline
$94460$ & ${>}26.3$ & $26.0$ & $24.6$ & $25.5$ & $25.7$ & $25.4$ & ${>}25.7$ & ${>}23.9$ & $25.2$ & ${>}23.4$ & ${>}24.8$ & ${>}24.5$ \\
$105937$ & $26.1$ & $26.1$ & $25.1$ & $25.5$ & $25.3$ & $25.2$ & $25.3$ & ${>}23.9$ & $25.0$ & ${>}23.4$ & $23.8$ & $23.9$ \\
$104037$ & $25.7$ & $24.8$ & $23.8$ & $24.3$ & $24.2$ & $24.2$ & $24.2$ & ${>}23.9$ & $24.1$ & $23.4$ & $23.8$ & $23.9$ \\
$100871$ & ${>}26.3$ & ${>}26.8$ & $25.6$ & ${>}26.9$ & ${>}26.9$ & ${>}26.6$ & ${>}25.7$ & ${>}23.9$ & $--$ & ${>}23.4$ & ${>}24.8$ & ${>}24.1$ \\
$101846$ & ${>}26.3$ & ${>}26.8$ & $25.9$ & ${>}26.9$ & ${>}26.9$ & ${>}26.6$ & ${>}25.7$ & ${>}23.9$ & $26.4$ & ${>}23.4$ & ${>}24.8$ & ${>}24.1$ \\
\hline
\end{tabular}

\\
\vspace{-1mm}
\begin{flushleft}
\scriptsize
Magnitude limits correspond to $3\sigma$ flux limits. Entries with ``$--$'' denote objects without F160W coverage.
\end{flushleft}
\end{table*}

\section{Results}
\label{sec:results}

We can now take full advantage of our large sample
with individually determined \fesc{} values and investigate possible
trends with other galaxy properties so we may better understand the mechanisms
through which LyC photons escape.


\subsection{Dependence on the Strength of \Lya}
\label{sec:fesc_strength_Lya}

The mechanisms through which LyC and \Lya{} photons escape may be very similar
if geometry plays a dominant role in the escape of ionizing photons. Low-density
channels created by a burst of star-formation may enable LyC and \Lya{}
photons to leak out in a specific direction. However, the radiative transfer of
LyC and \Lya{} differ, which may influence their relative visibility.
\Lya{} is absorbed by dust and is a resonant line and can thus
be scattered back into the line of sight by neutral hydrogen, whereas LyC photons
can only be absorbed by dust and neutral hydrogen. Therefore it may be possible
to have significant \Lya{} escape whilst that of Lyman continuum photons is
suppressed.

Contrary to this picture, extreme
\Lya{} emission would imply that most of the ionizing photons have been reprocessed as
\Lya{} photons, allowing few LyC photons to escape.
However, \cite{nakajima2014}
predict that significant \fesc{} and large \EWLya{} are possible
for ${f_{\rm esc}}<0.8$ and it is only for very extreme escape fractions
that \Lya{} will be suppressed.

Many authors have reported a positive correlation between
the \EWLya{} and escaping LyC photons.
\cite{verhamme2017} reported this empirical relation using the small
number of confirmed LyC-leaking low-redshift sources. At intermediate redshifts
the trend has also been observed using stacked spectroscopy
\citep{marchi2017} and ground-based imaging \citep{micheva2015}, the latter of
which may be affected by foreground-contamination.

\begin{figure*}
\centering
\includegraphics[width=\textwidth]{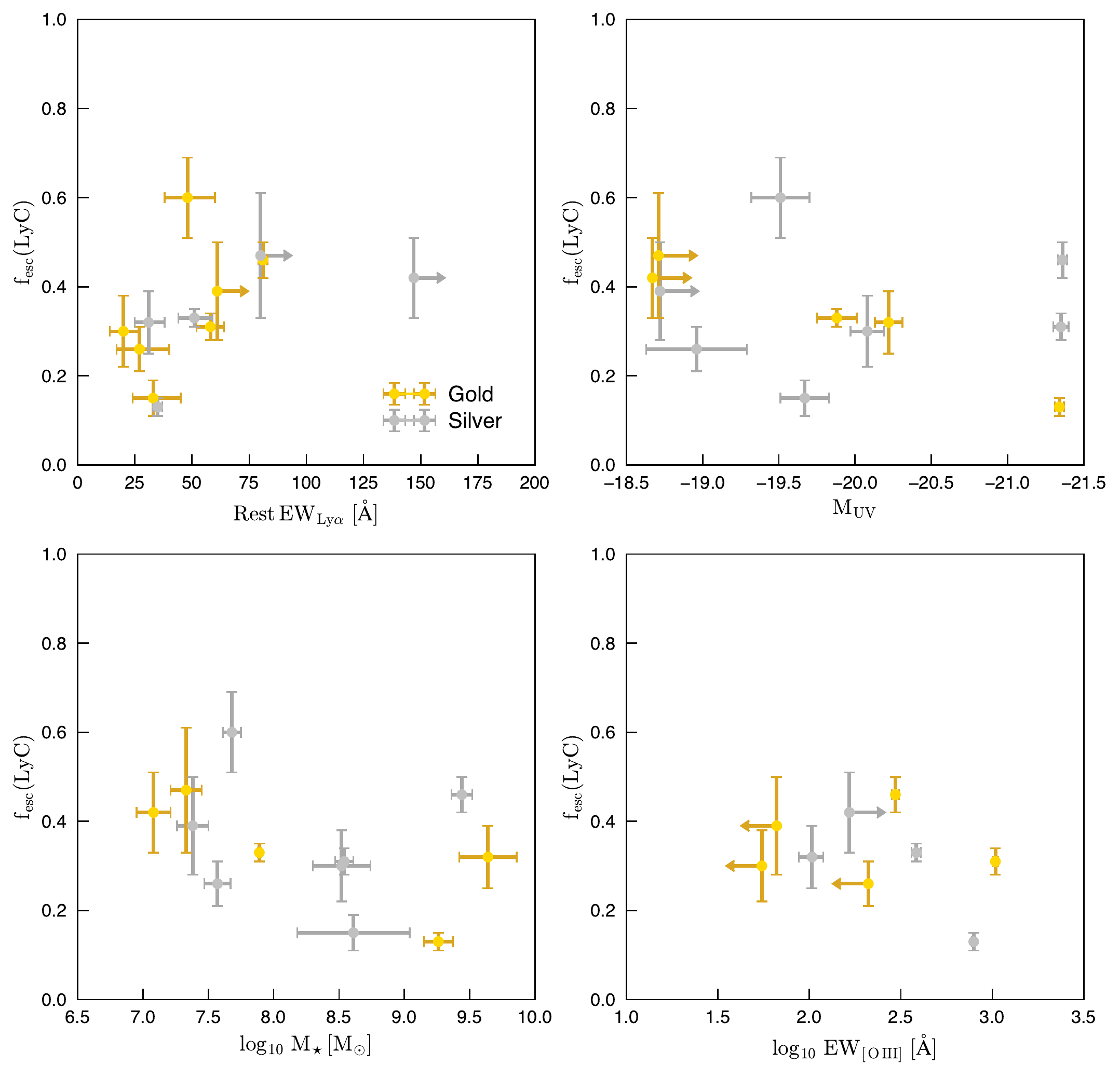}
\caption{Correlations between \fesc{} and other observed and derived properties
for the LACES Gold and Silver samples. Top-left: dependence of \fesc{} on \EWLya{}.
Top-right: dependence of \fesc{} on UV luminosity. Bottom-left: dependence of
\fesc{} on stellar mass. Bottom-right: Dependence of \fesc{} on the equivalent
width of \OIII{}. In the bottom-left panel both \fesc{} and stellar mass are
inferred model-dependent parameters where the stellar mass is the
product of the age and SFR derived from SED fitting. Indeed, at fixed age it is expected that
the SFR and \fesc{} would covary to match the $F336W$ measurements and could
produce this trend.}
\label{fig:correlations}
\end{figure*}

For the LACES sample deep imaging in the Subaru narrow-band covering all our
targets allows us to determine accurate estimates for the \EWLya{}. We use
the Subaru photometry instead of our spectroscopic data because the
extended \Lya{} flux can be included without slit losses and we can easily
compare the flux measured in the narrow band to the continuum measured in the broad
bands. We show our individually-determined escape fractions against \EWLya{} in
the top-left panel of Figure \ref{fig:correlations}. There does appear to be
a positive correlation albeit with large scatter.
Uncertainty in the likelihood of \fesc{} arises due to
degeneracies in choosing slightly different models to fit the SED of each
galaxy. It should be noted previous work exploring this correlation
at intermediate redshifts did not include such model-dependencies as the
observed flux density ratio of LyC to rest-frame UV photons was used instead of
\fesc{}.

We also calculate the escape fraction of \Lya{} (\fescLya)
by taking the ratio of the observed \Lya{} flux to the predicted intrinsic \Lya{}
flux derived from the observed $\rm H \beta$ flux using recombination physics,

\begin{equation}
	f_{\rm esc}^{\rm Ly\alpha} =
    \frac{F({\rm Ly\alpha})}{8.7 \times 2.86 \times F({\rm H\beta})},
\end{equation}
\noindent
where we use the ratios $\rm Ly\alpha/H\alpha = 8.7$ and $\rm H\alpha/H\beta =2.86$
\citep{hayes2015}.

A positive correlation between \fescLyc{} and \fescLya{} has been reported for the
small number of low-redshift LyC leaking galaxies \citep{verhamme2017}. The fact that
\Lya{} appears to escape preferentially to LyC may imply LyC escapes through channels in
a `riddled ionization-bounded nebula' \citep{zackrisson2013,behrens2014,verhamme2015},
whereas \Lya{} can escape additionally due to resonant scattering. Radiative transfer
simulations \citep{dijkstra2016} reproduce these trends albeit with much scatter
for \fescLya$>0.1$ due to the effects of dust, outflow kinematics and covering
factor. For the present sample, no clear correlation between \fescLyc{} and
\fescLya{} is seen, but uncertainties arising from our faint \Hb{} detections may mask a genuine trend.

At first sight, it is puzzling to find a relatively strong correlation between
\fescLyc{} and \EWLya{} but not between \fescLyc{} and \fescLya{} when $\rm EW_{Ly\alpha}
\propto f_{\rm esc}^{\rm Ly\alpha}\times(1-f_{\rm esc}^{\rm LyC})\times\xi_{\rm ion}$.
However, the \fescLyc{}-\EWLya{} correlation may have a marginal
dependence on \MUV{}. UV-fainter LACES objects tend to have larger \EWLya{}, as seen
for many samples in the literature \citep{stark2010,schenker2012,ono2012}. This may be
due in part to increased \xiion{} \citep{nakajima2018}. In Section \ref{sec:fesc_Mstar_MUV}
we show that \fescLyc{} is anti-correlated with UV luminosity and stellar mass.
Therefore, it is not surprising that LAEs with the most
extreme \fesc{} and \EWLya{} are intrinsically faint in \MUV{}.
Thus \fesc{} correlates with \EWLya{} because the fainter objects are more
compact with a harder \xiion{} which boosts \EWLya{}.


\subsection{Dependence of \fesc{} on Luminosity and Stellar Mass}
\label{sec:fesc_Mstar_MUV}

Understanding the typical \xiion{}, \fesc{} and \MUV{} of galaxies at $z>6$ is
crucial in determining whether galaxies were
the primary driver of cosmic reionization.
Current estimates assume an average \fesc{} and \xiion{}
and extrapolate the UV luminosity function down to a limiting magnitude, fainter
than current observations \citep{robertson2013}.
It has been
suggested that fainter galaxies may have higher \fesc{} or \xiion{}, contributing enough ionizing
photons such that galaxies alone are capable of reionizing the universe
\citep[e.g.,][]{inoue2006,kuhlen2012,bouwens2012a,finkelstein2012a,fontanot2012a,fontanot2014a,robertson2013,faisst2016}.

In \cite{nakajima2018} we have shown for a similar sample of LAEs
in the SSA22 protocluster that the production efficiency of ionizing photons,
\xiion{}, increases towards lower UV luminosities.
\cite{grazian2017} found tentative
evidence suggesting a trend between UV luminosity and
escape fraction using mostly limits on \fescrel{} derived from ground-based imaging.
However, all but 3 of their points were upper limits for \fescrel{}, therefore this result may
simply be due to the U-band imaging depth of their observations.
It has been suggested that such a correlation could arise from the fact that galaxies
with a lower luminosity will tend to also be lower mass and
there is an anti-correlation between stellar mass and the \Othreetwo{} ratio
\citep{nakajima2014,faisst2016,dijkstra2016} which itself is expected to correlate
with \fesc{} \citep{jaskot2013,nakajima2014,nakajima2016,faisst2016,izotov2018}.
Although, in the bottom-left panel of Figure \ref{fig:correlations}, we confirm that
the stellar mass is anti-correlated with \fesc{} for the same objects, this figure is
based on two SED-derived parameters, both of which are model-dependent.

The top-right plot in
Figure \ref{fig:correlations} shows the distribution
between \MUV{} and \fesc{} for the individual objects in the LACES Gold
and Silver samples. Among these sources, there appears to be no clear correlation between
UV luminosities and \fesc{} despite the correlation between stellar mass and \fesc.
 All the LACES objects
presented here are detected LyC leakers. There are a few outliers at
the bright end of our sample for which $f_{\rm esc} \sim 0.5$ although one of these
is an LAE-AGN.


\subsection{Dependence of \fesc{} on the Strength of \OIII{}}
\label{sec:fesc_OIII}

There is growing evidence that the \Othreetwo{} ratio correlates with \fesc{}
\citep{nakajima2014,nakajima2016,faisst2016,izotov2018}. Characterizing the mean
\fesc{} of $z\sim 3.1$ LAEs with extreme \Othreetwo{} will be useful
in understanding the role similar $z>6$ LAEs, where LyC emission is not directly
observable due to a partially neutral IGM, have in contributing to reionization.
Unfortunately, due to observational constraints,
we did not obtain deep enough \OII{} measurements to properly correlate
\Othreetwo{} with \fesc{} for a statistically meaningful sample.

In the absence of \OII{} measurements we instead use the equivalent
width of \OIII{} as a large \EWOIII{} may imply a large
\Othreetwo{} ratio. This appears to be a reasonable assumption and \citep{tang2018a} find
that galaxies with large \EWOIII{} almost always have large \Othreetwo{}. Moreover, it appears that galaxies in the reionization-era
differ in that they have more extreme \OIII{} as the strength of the \OIII{} line
appears to increase
with redshift \citep{schenker2013,smit2014,smit2015}. The discovery that LBGs at $z>7$ with
extreme \OIII{} have been detected with \Lya{} emission also implies a large \fesc{} as these
objects may have ionized
bubbles of hydrogen early so that their \Lya{} emission could
redshift out of resonance with neutral hydrogen and escape
\citep{guido2016,zitrin2015,oesch2015,laporte2017,stark2017}.
This assumption appears reasonable  as \citet{tang2018a} find that galaxies with large \EWOIII{} almost always have large O32 and large \xiion{}.

We use the equivalent width instead of the flux, as our
objects span a wide range of magnitudes. Therefore, in order to accurately calculate
the \EWOIII{} we require that our LAEs were targeted and detected in
our MOSFIRE campaign but also that our targets were covered by the deep HST $F160W$
photometry so we can accurately estimate the continuum at the \OIII{} line (see
section \ref{sec:NIR_spectroscopy}).

We show the results for the LACES sample in the bottom-right
panel of Figure \ref{fig:correlations}.
There is no clear correlation
between \EWOIII{} and \fesc{} although the scatter is large. Accordingly, we cannot test the
physically-motivated hypothesis that density-bounded nebulae result in LyC leakage.
Further \OII{} measurements would enable us to
correlate \fesc{} directly with \Othreetwo{} for the LACES Gold and Silver
subsamples. We will present results from further Keck/MOSFIRE observations of our
sample in future work (Nakajima et al., in preparation).


\section{Discussion}
\label{sec:discussion}

The robust detection of Lyman continuum photons from a substantial subset
of our LAE sample, combined with stringent limits on Lyman continuum escape
in our non-detected objects, may provide clues as to how galaxies physically
release hydrogen ionizing photons as required if they drove cosmic reionization.
Below, we examine possible differences between our detected and non-detected
samples, discuss possible physical mechanisms for the escape of Lyman continuum
photons that may explain our results, and compare with previous searches for
Lyman continuum emission in galaxies.


\subsection{Understanding the Non-Detections}
\label{sec:understanding_non_detections}

We now discuss the puzzling dichotomy between our LyC detections and non-detections.
The 12 LyC-leaking LAEs are all individually detected with $\rm SNR>4$ in the $F336W$ images
and have $f_{\rm esc} \sim 15-60 \%$, whereas even in a mean composite spectrum of $38$
non-detections we estimate $f_{\rm esc} <0.5 \%$.

\begin{figure*}
\centering
\includegraphics[width=\textwidth]{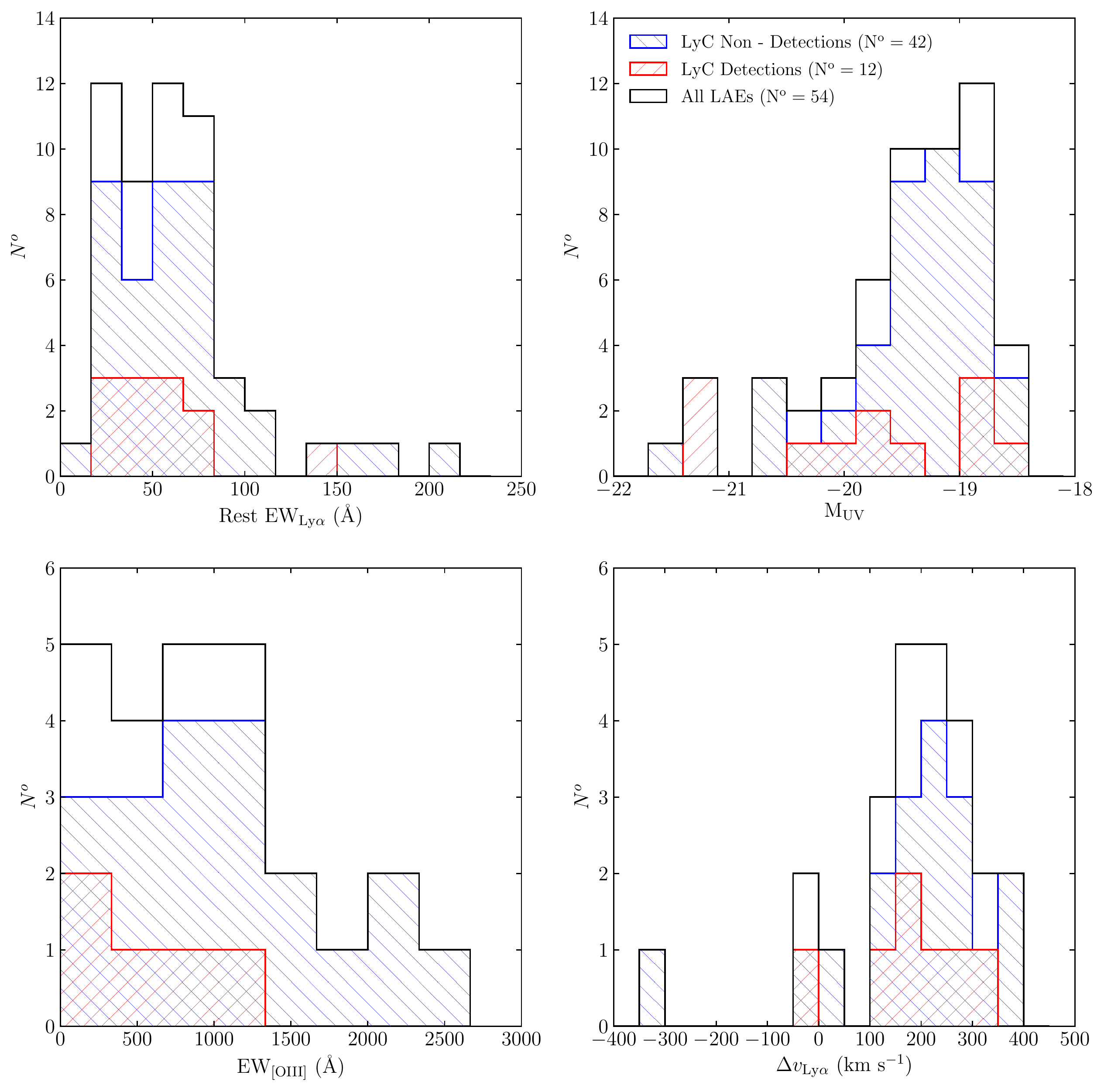}
\caption{Distributions of \EWLya{} (top-left), \MUV{} (top-right), \EWOIII{} (bottom-left) and \dvLya{} (bottom-right) for the LACES sample. Red hatched histograms show the numbers of LAEs with detected $F336W$ emission and blue histograms show non-detected LAEs. The black outline shows the total number of LAEs in the detections and non-detections. The \OIII{} line and both a \Lya{} and systemic redshift were not always targeted or detected in our extensive optical and near-infrared spectroscopy. Therefore, in the two lower panels showing the distributions of \EWOIII{} and \dvLya{} we only show LAEs for which the relevant data is complete.}
\label{fig:fesc_populations}
\end{figure*}

We first investigate whether there are any differences between these two populations
in terms of their luminosity (\MUV{}), strength of \Lya{} emission (\EWLya{}), \EWOIII{} and velocity offset of \Lya{} (\dvLya{}).
In Figure \ref{fig:fesc_populations} we show how the detections, including Gold and Silver subsamples, and the non-detections ($N=42$) are distributed
across these parameters with reference to the full sample of $54$ LAEs. It appears that the population of detections and non-detections are almost
indistinguishable from one another. LyC-detected objects span the full range of UV luminosities as do the non-detections.
However, it should be noted that there is an observational bias at fainter
luminosities as we will not be able to detect small escape fractions in individual cases
for the faintest LAEs and it remains possible that some of the faint LAEs in the
LyC non-detections are weak to moderate LyC-leakers but below our detection limit.
Nevertheless, if this were the case we would still expect to detect this faint signal in the
deep $F336W$ stack, yet we recover on average $f_{\rm esc} <0.5 \%$.

Additionally, LyC-detected LAEs have a similar distribution in \EWLya{} with respect
to non-detections. However, our LyC detections are almost all at
$\rm EW_{Ly\alpha}< 100 \,\AA$,
as is the case for most of the LAEs in the LACES sample.
Of the $6$ LAEs with $\rm EW_{Ly\alpha}> 100 \, \AA$ only $1$ is a LyC leaker.
This trend may be expected if
LAEs with very large \EWLya{} have reprocessed almost all of
their ionizing photons into \Lya{}, resulting in
galaxies with $f_{\rm esc} \sim 0$
(e.g., \citealt{nakajima2014}).
These LAEs are also the faintest objects discussed above,
so even if they have a non-zero \fesc{} it will most likely be below our detection limit.
Regardless, the bulk of the non-leaking LAEs contributing to the composite spectrum
have moderate \EWLya{} and \MUV{} and fall in the same space in a \EWLya{}-\MUV{} plot
as the bulk of the detections. Therefore, given the evidence from the detections
(Section \ref{sec:fesc_strength_Lya} and \ref{sec:fesc_Mstar_MUV}) we
would have expected these LyC non-detections to have moderate \fesc{}
that would be detectable in our deep $F336W$ stack.
Yet, we detect no signal when stacking these objects and find $f_{\rm esc} < 0.5 \%$. This could be due to differences in the covering factor of these LAEs, however
we do not probe this property in our observations.

We also examine the distribution of \EWOIII{} in Figure \ref{fig:fesc_populations}.
Once again, the detections cover the full range of \EWOIII{} as do the non-detections.
If LyC leakage arises due to density-bound nebulae with extreme \Othreetwo{} ratios
this may imply extreme \EWOIII{} \citep{tang2018a}. We may therefore expect the LyC detections to be preferentially clustered at large \EWOIII{} compared to the non-detections, but
the detections fall at \EWOIII{}$\lesssim 1400 \rm \AA$.
Again, it may be the case that if galaxies are leaking a significant fraction of their ionizing photons there are few $>35 \, \rm eV$ photons remaining to doubly ionize oxygen. This could perhaps explain why some of our LyC leakers have smaller \EWOIII{}. It is important to note that \EWOIII{} measurements have not been obtained for all the LAEs as \OIII{} was not targeted or detected for every object and we do not have full $F160W$ coverage for the LACES sample in order to estimate the continuum at $\sim 5000$ \AA{} in the rest-frame.

Finally we examine the distributions of \dvLya{}, which should be $< 150 \, \rm km \, s^{-1}$ for LyC leakers that require a low column density of neutral gas for LyC escape \citep{verhamme2015}. We find no significant difference between detections and non-detections. We will discuss this result in more detail below in Section \ref{sec:anisotropic}.

To summarize $\sim 20 \%$ of the LAEs in the LACES sample have individual $F336W$ detections with
inferred escape fractions ranging
from $15 \%$ to $60 \%$. Using composite SEDs we find the average
$f_{\rm esc} = 0.20\pm0.02 \%$ and $f_{\rm esc} = 0.51\pm0.08 \%$
for the Gold and Silver samples respectively. Even when using a stack of 32 LAEs
that are not detected as LyC-leakers in individual $F336W$ images we find on average
$f_{\rm esc} < 0.5 \%$ despite these LAEs having an almost identical distribution of
luminosities, \EWLya{}, \EWOIII{} and \dvLya{} as the LyC-leakers. We now turn to possible explanations
for this dichotomy.

\subsubsection{Anisotropic LyC Escape}
\label{sec:anisotropic}

Previous analyses have proposed \citep{zackrisson2013,behrens2014,nakajima2014}
that there exist at least two mechanisms through which
LyC photons can escape. The first involves
an ionization-bounded nebula where \HII{} and \HI{} shells surround the central
stars. LyC photons can escape from such a system
if stellar winds or supernovae from a burst of star-formation produce low-density holes
through the neutral hydrogen in the ISM. LyC photons can then easily escape through
these channels without being absorbed. Alternatively if the stellar population has a very hard
spectrum or there is a significant burst of star-formation,
the resulting ionization of the gas may enable
LyC photons to readily escape in all directions
\citep{zackrisson2013,behrens2014,nakajima2014}.

If LAEs are ionization-bound with holes then it will only be possible to detect LyC
leakage if our line of sight is coincident with the opening angle of these channels. All LAEs
in a given sample could be leaking LyC radiation
but only a fraction of them, corresponding to the average covering fraction of \HI{} and
dust, would be detected as LyC leakers through direct observations. The angular
dependence of the escape fraction is found to be highly anisotropic in simulations, with
galaxies with smaller \fesc{} having a smaller solid angle through which LyC photons can
escape \citep{paardekooper2015}. Therefore, if the escape of LyC photons occurs
anisotropically through channels, it seems likely that our non-detections would
have small \fesc{}, with the photons escaping out of small channels directed away from
our line of sight. However, we do not probe the covering fraction of our LAEs and we
cannot be certain that geometric effects are the main cause for the dichotomy between
our detections and non-detections.

Using \Lya{} transfer calculations in \HI{} regions, \cite{verhamme2015} showed that if LyC
escapes due to an optically thin ($\rm N_{HI} \leq 10^{18} \, \rm cm^{-2}$),
density-bounded regime then the \Lya{} profile will be narrow with a small velocity offset
($\Delta v_{\rm Ly\alpha} < 150 \, \rm km \, s^{-1}$). However, if the LyC-leakers are
ionization-bounded and riddled with low-density channels
$\Delta v_{\rm Ly\alpha} \sim 0 \, \rm km \, s^{-1}$ with a small red peak due to additional
scattered \Lya{} light that then escapes through the channel. If the dichotomy between our
detections and non-detections is caused by geometry we might expect the LyC-leakers to be
preferentially clustered around $\Delta v_{\rm Ly\alpha} \sim 0 \, \rm km \, s^{-1}$ when
compared to the non-detections.

In Figure \ref{fig:fesc_populations} we show the distribution of
\dvLya{} for the detections and non-detections where both a \Lya{} and
systemic redshift are available. The LyC detections cover the full range of velocity offsets
and are not centered only around small velocity offsets. Therefore it is not clear
that the non-detections are ionization-bounded whereas detections are riddled or density-bounded. Indeed, the \Lya{} profile is most likely more complex than this
simple picture. Velocity offsets could be caused by outflows. Also, if LyC escapes
through small offset channels, this geometry could have little implication for the \Lya{}
profile, which may still be dominated by resonant scattering.
More detailed analysis of \Lya{} spectra at higher spectral resolution are most likely needed
before ruling out the geometric picture of \Lya{} escape.


\subsubsection{Stochastic LyC Escape}
\label{sec:stochastic}

Star formation at these redshifts may be highly time-dependent. Galaxies accrete gas from the IGM and through mergers. They undergo bursts of star-formation, as a result of which feedback in the form of
stellar winds and supernovae can ionize their ISM. During the relatively quiescent periods, the ionized
gas will recombine. LyC leakage may therefore be stochastic with bursts of star-formation either ionizing all the neutral hydrogen within the virial radius creating channels through which LyC photons can escape. This has been widely reported in simulations of leaking LyC radiation where \fesc{} has
traced bursty star formation with a time delay of $\sim 10$ Myr
\citep{kimm2013,wise2014,ma2016,kimm2016,trebitsch2017}
and with smaller, lower mass galaxies expected to be more stochastic.

We would therefore expect that the LACES LAEs with no leaking LyC radiation are being observed in these quiescent periods where the ISM has had time to recombine. However, this
picture is difficult to reconcile due to large \EWOIII{} we observe for the non-detections,
implying very recent star-formation. These galaxies may be recently star-forming but feedback may not have been effective in creating pathways for the
radiation to escape. This ineffectiveness may trace an additional factor such as the
covering fraction which could be varying between individual galaxies.
Thus, galaxies with lower covering fractions are more able
to leak LyC, given the same burst of star-formation and feedback.

In this scenario, only a fraction of LAEs at any time
would be at a point where they had recently undergone a burst of star-formation $\sim 10$
Myr ago and only these LAEs would be detectable as having a non-zero \fesc{}. If it were
possible to observe these LAEs over hundreds of Myr perhaps we would see the LAEs
flash ``on'' and ``off'' in LyC emission.


\subsubsection{Spatially Varying Intergalactic Medium}
\label{sec:varying_IGM}

As our study takes place within the SSA22 protocluster, a further possibility
is that the IGM is spatially varying on scales of tens to a few hundred Mpc within our
area of study and the field of view of the $F336W$ images. This density
variation would result in LyC
leakage from LAEs in regions with a higher column density of \HI{} being more
strongly absorbed and
thus we would measure a reduced or zero \fesc{} for these objects.

Evidence for spatial variation of \HI{} in the IGM and CGM within the SSA22 protocluster has
been investigated in the literature. Using bright galaxies behind the
protocluster, \cite{mawatari2017} measured \Lya{} absorption in the spectrum of these
galaxies due to absorption by \HI{} within SSA22.
They found that SSA22 has an excess of
\HI{} compared to similar independent control fields. They also
found that there is a large scale diffuse \HI{} component which is independent of the
CGM of individual galaxies, \Lya{} absorption increased on $<100 \, \rm Mpc$ scales possibly
due to the CGM of nearby galaxies and that stronger LAEs had weaker \HI{} absorption.

\begin{figure}
\centering
\includegraphics[width=.5\textwidth]{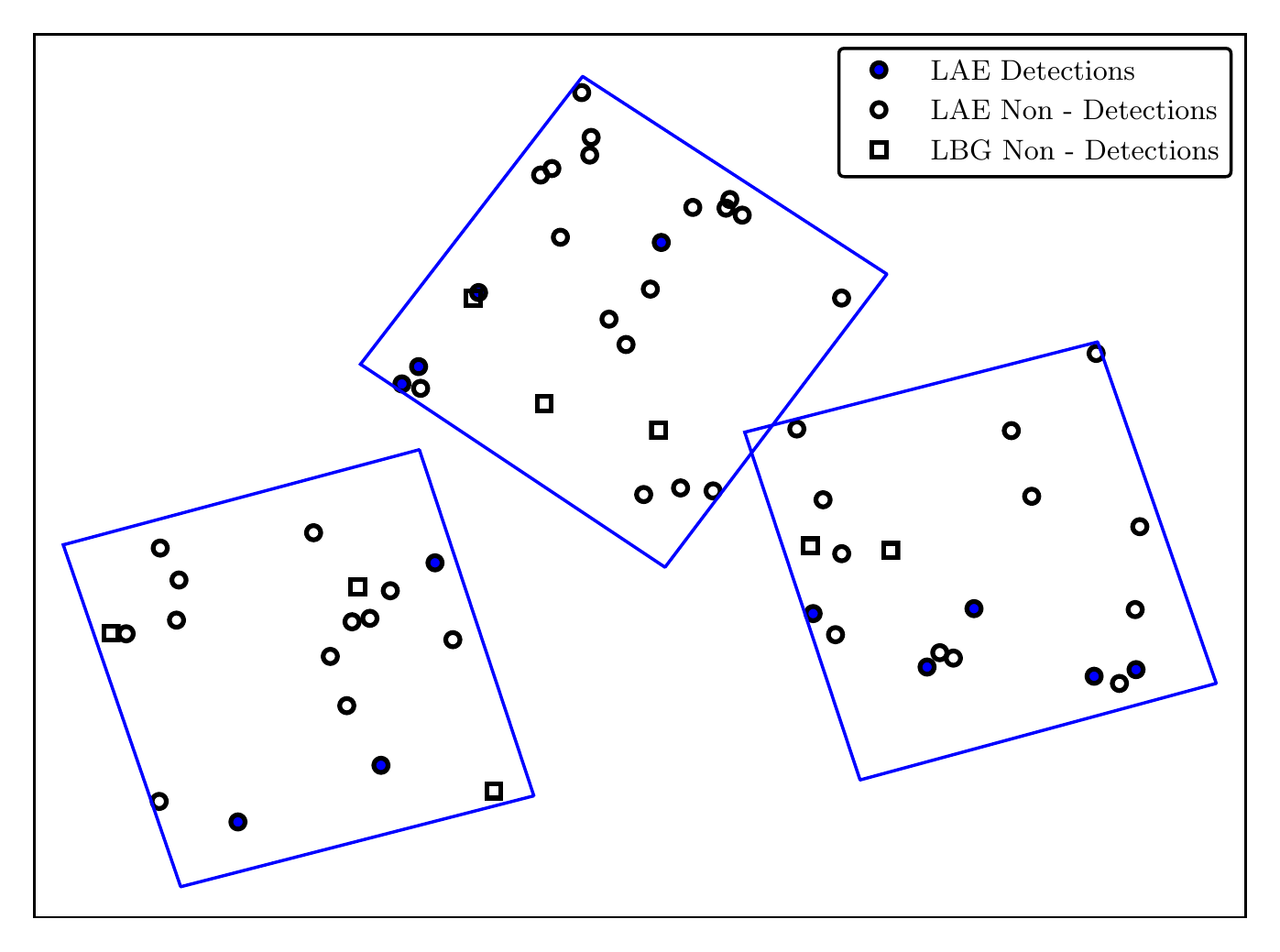}
\caption{Spatial distribution of LyC detections (blue) and
non-detections (white) for all LAEs and LBGs (circles and squares respectively)
in the LACES sample.
Detections and non-detections appear to be spatially clustered which could be
due to spatial variations of \HI{} gas in the SSA22 protocluster.}
\label{fig:spatial_dist}
\end{figure}

\cite{mawatari2017} focused on the center of SSA22 and the LACES
sample is drawn from the edge of the region considered in their study. It is therefore possible that
there is large scale diffuse \HI{} unconnected to individual LAEs within our field
and that a particularly dense CGM of a nearby galaxy or an ionizing neighbor with a large
\fesc{} may respectively inhibit or boost the chances of leaking LyC radiation reaching us
as observers. Figure \ref{fig:spatial_dist} shows the spatial distribution of
LAE-LyC detections (blue circles) and LyC-non-detections for LAEs and LBGs (white circles
and squares respectively).

Indeed, it does seem that the LyC detections and non-detections
appear clustered on small scales which could be due to spatially varying \HI{} in the
IGM and CGM. We cannot directly measure the \HI{} in the LACES field but future detections
of leaking LyC radiation in clusters with lower \HI{} density or in blank fields may
help us understand if LyC non-detections in extreme LAEs owes to \HI{} absorption. It should be noted that this effect would only change the fraction of leaking LyC radiation that is absorbed along the line of sight. We have still detected LyC escape in a significant fraction of our LAEs but it is possible more IGM absorption results in non-detections for the remaining LAEs.

\subsection{Comparison to Other Studies}
\label{sec:compare_studies}

Through the LACES program, we have provided a significant number of individual LyC measurements for a homogeneous sample of star-forming galaxies in a narrow redshift interval at $z\simeq 3.1$, corresponding to escape fractions of \fesc{} $\simeq 15-60 \%$. Of particular significance is the high success rate ($\simeq 20 \%$) within our sample, one considered to be potential analogs of \OIII{}-strong metal-poor sources in the reionization era.

Early efforts to directly measure significant escape fractions in galaxies at intermediate redshifts have largely been unproductive apart from a few exceptional cases \citep{shapley2016,barros2016,vanzella2016,bian2017,vanzella2018}.
Without a sizable number of detections drawn from a homogeneous sample, it has therefore been difficult to make progress in understanding under what conditions LyC photons can escape.

An alternative approach when no individual detections can be found, for example due to the limited sensitivity of the data, is to stack the LyC signal from a large sample either with suitable photometry
\citep{rutkowski2017,matthee2017,grazian2017,japelj2017,naidu2018} or spectroscopic data \citep{marchi2017,steidel2018}. However, with the exception of the recent campaign by \citet{steidel2018}, this has resulted primarily in upper limits of $ f_{\rm esc} < 10 \%$ below the
canonical value of $10-20\%$ required to sustain reionization \citep{robertson2013}. The results of these programs has led to speculation that reionization may not be driven by star-forming galaxies e.g. \cite{madau2015}.

An important conclusion from our work, which would not be easily seen in early stacking programs, is the distinction between the 12 LACES LAEs which show convincing individual detections in the range $f_{\rm esc} \simeq 15-60 \%$ and 42 LAEs which, even when stacked, show no significant leakage consistent with an individual average $f_{\rm esc} < 0.5 \%$. If, as we surmise, LyC leakage is either ``on" or ``off'' due to anisotropic or time-varying factors, then {\it achieving adequate depth for an individual target is crucial to making the distinction}. If $z>7$ sources had the same inferred escape fraction as our LyC detections, they could maintain reionization. Yet shallower surveys of sources less analogous to $z>7$ objects would have concluded the opposite.

As an illustrative example, stacking $F275W$ photometry \cite{rutkowski2017} found $1\sigma$ upper limits of $f_{\rm esc}<14.0\%$ for $13$ `extreme emission line galaxies' (EELGs) at $z\sim 2.3$ with $\rm O_{32} > 5$. Although $3\sigma$ upper limits derived from their measurements could still be consistent with significant \fesc{} this result casts doubt on whether galaxies with
extreme \Othreetwo{} are LyC leakers. Similarly \cite{naidu2018} find $f_{\rm esc}<16.7\%$ for fainter EELGs and $f_{\rm esc}<8.5\%$ for their brighter EELGs using stacked ground-based U band imaging. However, if only a fraction of the LyC photons are escaping in our direction, then they will remain
undetected in the relatively shallow images used by \cite{rutkowski2017} ($3\sigma$ depth ranging $26.5-28.2$ AB) whose depth was optimized for a composite stack.

The \citet{steidel2018} spectroscopic campaign is the only comparative study which reaches a depth adequate for individual LyC detections comparable to escape fractions of $\simeq$10 percent. Individual detections were seen for 15/124 sources. The approach is highly complementary to the present study in several respects. It focuses on Lyman break galaxies (LBGs) over a wider redshift range with LyC signals inferred optimally from spectra in a narrow wavelength window (880-910 nm) where IGM absorption is reduced and samples multiple sight-lines to reduce cosmic variance. The LACES program avoids some of the limitations of earlier HST imaging campaigns which targeted LBGs with a range of redshifts. LACES exploits a narrow-band selected sample of LAEs at $z\simeq 3.1$ optimally matched to the F336W filter and the improved depth of the HST imaging (a $3\sigma$ limiting magnitude of $\simeq 30.2$) to provide exquisite limits on individual sources with the necessary resolution to mitigate issues of foreground contamination.

Comparing the two approaches, the success rate of detecting LyC emission in the
LACES sample ($\rm SNR\geq 4$ detections for $20 \%$ of the total sample and $21 \%$ for LAEs only) is higher than that seen for individual LBGs ($\simeq$10\% to broadly comparable escape fractions in the \citealt{steidel2018} survey). Indeed, none of the 7 LBGs in our LACES control sample have detectable LyC emission. Our earlier work has shown LAEs have a harder \xiion{} than LBGs \citep{nakajima2016,nakajima2018}. With more ionizing photons, significant LyC leakage is more likely for LAEs and also results in larger \Othreetwo{} ratios and more extreme \OIII{} equivalent widths. LAEs can more readily leak LyC photons in riddled ionization-bounded or density-bounded nebulae, physical conditions more easily met for younger, low mass and metal-poor galaxies with important implications for comparable sources in the reionization era.

Despite the different approaches, many of the conclusions of the present paper are supported by the \citet{steidel2018} results including the absence of any demographic differences between the sample of individual detections and those non-detected and similar correlations between \fesc{} and \EWLya{} and \MUV{}. The primary difference remains the higher success rate of detecting LyC leakages in LAEs and the possible association with \OIII{} emission.

\subsection{Implications for Cosmic Reionization}

Our interest in the LACES sample and this study is motivated,
in part, by the likelihood that our $z\simeq 3.1$ LAEs are promising
analogs of sources in the reionization era and thus that inferences
on the physical conditions that permit LyC photons to escape will
have important implications for the assumption that cosmic reionization
is largely driven by similar systems.

It is reasonable to assume that LAEs at intermediate redshifts, that are metal-poor,
low mass star-forming galaxies are similar to those at higher redshift. However, further
simularities between our sample and typical $z>7$ galaxies is based on meager data
at high redshifts. These include promising indications that $z>7$ galaxies
have hard ionizing spectra
\citep{stark2015,stark2017,laporte2017,mainali2018} as seen in the
LACES sample \citep{nakajima2016,nakajima2018}, as well as intense
\OIII{} emission characteristic of many high redshift IRAC-excess sources
\citep{smit2014,smit2015,guido2016}.
Assuming this is the case, what can be deduced from the fact that
approximately $\simeq 20\%$ of our LACES LAEs meet the
canonical criterion for an escape fraction $f_{\rm esc}\geq 10\%$ required
to drive reionization \citep{robertson2013,brant2015constraints}?

At face value, the {\it average} escape fraction for our LACES sample
is substantially diminished by the dominant population
for which no significant LyC leakage was detected to quite
impressive limits. Nominally the average $f_{esc}$ would be reduced from
20\% (the mean of the Gold and Silver samples) to only 5\% - a figure
in reasonable agreement with the value determined for LBGs by
\citet{steidel2018}.

However if, as seems possible, the dichotomy between our
detections and non-detections is largely due to anisotropic
LyC leakage, it is likely the majority of the LACES LAE are significantly
influencing their local IGM. If a similar behavior is present
in the reionization era, then such a coarse average would underestimate
the role that early equivalents of the LACES population would play
in governing reionization. Less luminous versions of our LAEs
at high redshift could well have even higher escape fractions, as
hinted by the trends in Figure 13. Additionally, the intrinsic fraction of LAEs
is observed to increase with redshift \citep{stark2010,schenker2012} and so
we expect that an increasingly large fraction of high-redshift star-forming
galaxies will look more and more like the LACES sample.
Clearly, then, it is crucial to
understand physically the dichotomy discussed in Section 6.1.

With this in mind, in later papers we will explore the dependence of $f_{esc}$
on the ratio of \OIII{}/\OII{} to test the density-bound concept
first promoted by \citet{nakajima2014}. The present MOSFIRE
spectroscopic data has inadequate coverage of \OII{} emission
so such correlations cannot yet be examined. In addition, if
the dichotomy discussed above originates via anisotropic
LyC leakage, numerical simulations suggest that such high
escape fractions may arise when feedback creates a turbulent
interstellar gas enabling leakage through porous low density channels
\citep{kimm2013,wise2014}. This can be readily tested via
IFU spectroscopy which aims to correlate our HST-determined
escape fractions with spatially resolved ISM kinematics.


\section{Summary}
\label{sec:summary}

We present the first results from the LymAn Continuum Escape Survey (LACES),
where we obtained
ultra-deep HST WFC3 UVIS/$F336W$ imaging of a sample of 61 faint $z\simeq 3.1$ LAEs and LBGs in the
SSA22 field. The extreme depth of the $F336W$ images enabled individual direct detection of
escaping Lyman continuum emission ($\rm SNR\geq 4$) in 12 LAEs ($20 \%$) in our homogeneous sample.
Our program provides a huge increase in the number of individually detected
LyC leakers at intermediate redshift and represents the first time a large sample of LAEs
with a significant fraction of individual leakers has been presented.
We make use of extensive
multi-band photometry, including newly obtained HST WFC3 IR/$F160W$ imaging,
to fit the SED of each galaxy to obtain accurate individual
estimates of the escape fraction \fesc{}. We further use our SED fitting method to infer typical escape
fractions from composites of various subsamples.
For individual objects we obtain $f_{\rm esc} \approx 15-60 \%$
and for composites of our Gold and Silver subsamples of Lyman continuum detected objects
we find $f_{\mathrm{esc}}\geq20\%$. For our composite of the Lyman continuum
non-detection subsample, we infer $f_{\mathrm{esc}}\lesssim0.5\%$.

We expect the rate of contamination to be low ($98\%$ probability $\leq 2$ of our detections could be contaminants) and we ensure against foreground interlopers
using the high spatial resolution provided by the $F336W$ and $F160W$ images. We find that the escape fraction may
increase for low stellar-mass galaxies with larger \EWLya{}.

We discuss the dichotomy between our detections with significant \fesc{} and our non-detections,
seemingly drawn from the same sample, covering the same range of UV luminosities, \EWLya, \EWOIII{} and \dvLya{}. We suggest the reason for this dichotomy could owe to three factors: anisotropic
escape where LyC photons escape through channels and the opening angle of these
channels is only aligned with our line of sight for the detections, a time varying \fesc{} due
to the bursty nature of star-formation in these low-mass systems, or spatially varying \HI{}
in the IGM of the SSA22 protocluster which appears to have a higher \HI{} density compared to similar
control fields \citep{mawatari2017}.

The large \fesc{} detected by the LACES program, the fraction of LAEs which leak LyC photons,
and the large difference in Lyman continuum flux between detected and non-detected objects
may hold significant implications for understanding the mechanisms through which hydrogen
ionizing radiation escapes from galaxies. Coupled with our observations that suggest
low-mass LAEs with strong \Lya{} have the most extreme \fesc{},
our results provide exciting hints for how to answer the question of
whether galaxies served as the primary driver of cosmic reionization.\\


\acknowledgements{We thank the anonymous referee for helpful comments that improved our
manuscript.
We acknowledge financial support from European Research Council Advanced Grant FP7/669253 (TF, RSE).
BER acknowledges a Maureen and John Hendricks Visiting Professorship at the Institute for Advanced Study,
NASA program HST-GO-14747, contract NNG16PJ25C, and grant 17-ATP17-0034. KN acknowledges a JSPS Overseas Research Fellowship
and a JSPS Research Fellowship for Young Scientists.
DPS acknowledges support from the National Science Foundation through grant AST-1410155.
AKI is supported by the Japan Society for the Promotion of Science, KAKENHI Grant Number 17H01114.
This work is based on observations taken by the NASA/ESA HST, which is operated by the Association of Universities for Research in Astronomy, Inc., under NASA contract NAS5-26555.
We thank Masato Onodera for observing 1 of our 4 MOSFIRE K band masks. We also thank
T. Hayashino, T. Yamada and Y. Matsuda for providing
the LAE catalogue and the ground-based photometric data. We would also like to thank K. Kakiichi and N. Laporte for useful comments. Further data was taken with the Subaru telescope and the W.M. Keck Observatory on Maunakea, Hawaii the latter of which is operated as a scientific partnership among the California Institute of Technology, the University of California and the National Aeronautics and Space Administration. This Observatory was made possible by the generous financial support of the W. M. Keck Foundation. The authors wish to recognize and acknowledge the very significant cultural role and reverence that the summit of Maunakea has always had within the indigenous Hawaiian community. We are most fortunate to have the opportunity to conduct observations from this mountain.}


\bibliographystyle{aasjournal}  
\bibliography{main} 

\end{document}